\documentclass[aps]{revtex4-2}%
\usepackage{amsfonts}
\usepackage{amsmath}
\usepackage{amssymb}
\usepackage[colorlinks,linkcolor=blue,citecolor=blue,urlcolor=blue]{hyperref}
\usepackage{subfigure}
\usepackage{graphicx}
\usepackage{cleveref}
\usepackage{listings}
\usepackage{xcolor}
\usepackage{array}
\usepackage{url}%
\begin{document}
	\preprint{CTP-SCU/2020024}
	\title{Thermodynamic Geometry of Black Holes Enclosed by a Cavity in Extended Phase Space}
	\author{Peng Wang}
	\email{pengw@scu.edu.cn}
	\author{Feiyu Yao}
	\email{yaofeiyu@stu.scu.edu.cn}
	\affiliation{Center for Theoretical Physics, College of Physics, Sichuan University,
		Chengdu, 610064, China}

\begin{abstract}
Recently, the phase space of black holes in a spherical cavity of radius
$r_{B}$ has been extended by introducing a thermodynamic volume $V\equiv4\pi
r_{B}^{3}/3$. In the extended phase space, we consider the thermodynamic
geometry, which provides a powerful tool to understand the microscopic
structure of black holes, of Reissner-Nordstr\"{o}m (RN) black holes in a
cavity, as well as that of Reissner-Nordstr\"{o}m-AdS black holes. Although
the phase structures of the cavity and AdS cases show striking resemblance, we
find that there exist significant differences between the thermodynamic
geometries of these two cases. In particular, a reentrant transition of the
type of the microstructure interactions, i.e., repulsive $\rightarrow$
attractive $\rightarrow$ repulsive with increasing temperature in an isobaric
process, is observed for RN black holes in a cavity.

\end{abstract}

\keywords{}\maketitle
\tableofcontents

\section{Introduction}

\label{Sec:IN}

Black hole thermodynamics has been of crucial importance in our understanding
of quantum gravity. Since classical black holes absorb all matter and emit
nothing, black holes were not considered to be thermodynamic systems until the
Hawking's area theorem was proposed \cite{Hawking:1971tu}. Bekenstein
subsequently noticed a possible relation between the area theorem and the
second law of thermodynamics \cite{Bekenstein:1972tm}. The discovery of the
Hawking radiation then assigns black holes a temperature
\cite{Hawking:1974rv,Hawking:1975iha}, which further confirms the analogy
between usual thermodynamics and black hole mechanics. Consequently, the
\textquotedblleft four laws of black hole mechanics\textquotedblright\ were
proposed and formulated by Bardeen, Carter, and Hawking \cite{Bardeen:1973gs}.
Since the advent of the AdS/CFT correspondence
\cite{Maldacena:1997re,Gubser:1998bc,Witten:1998qj}, of particular interest
are asymptotically anti--de Sitter (AdS) black holes, which can be
thermodynamically stable as AdS boundary plays a role of a reflecting wall.
Moreover, the first law becomes consistent with the corresponding Smarr
relation, and black hole mass is interpreted as a chemical enthalpy
\cite{Kastor:2009wy}. After the Hawking-Page phase transition was observed in
Schwarzschild-AdS black holes \cite{Hawking:1982dh}, thermodynamic properties
and phase behavior of various more complicated black holes have been studied
\cite{Witten:1998zw,Cvetic:1999ne,Chamblin:1999tk,Chamblin:1999hg,Caldarelli:1999xj,Cai:2001dz,Cvetic:2001bk,Nojiri:2001aj,Wei:2012ui,Gunasekaran:2012dq,Cai:2013qga,Xu:2014kwa,Frassino:2014pha,Dehghani:2014caa,Hennigar:2015esa,Caceres:2015vsa,Wei:2015iwa,Hendi:2016yof,Hendi:2017fxp,Lemos:2018cfd,Pedraza:2018eey,Wang:2018xdz,Wei:2019uqg,Wei:2020poh,Guo:2021ere}%
.

On the other hand, York first reported that a Schwarzschild black hole
enclosed by a cavity can be thermally stable and also experiences a
Hawking-Page-like transition to the thermal flat space as the temperature
decreases \cite{York:1986it}. Later, thermodynamics of Reissner-Nordstr\"{o}m
(RN) black holes in a cavity was discussed in a grand canonical ensemble
\cite{Braden:1990hw} and a canonical ensemble
\cite{Carlip:2003ne,Lundgren:2006kt}. Like RN-AdS black holes, it showed that
a Hawking-Page-like phase transition occurs in the grand canonical ensemble,
and a van der Waals-like phase transition occurs in the canonical ensemble.
Moreover, Hawking-Page-like or van der Waals-like phase transitions have also
been found in the brane systems
\cite{Lu:2010xt,Wu:2011yu,Lu:2012rm,Lu:2013nt,Zhou:2015yxa,Xiao:2015bha}.
Properties of boson stars and hairy black holes in a cavity were also
investigated
\cite{Sanchis-Gual:2015lje,Sanchis-Gual:2016tcm,Basu:2016srp,Peng:2017gss,Peng:2017squ,Peng:2018abh}%
, which were shown to closely resemble those of holographic superconductors in
the AdS gravity. Later, the authors of \cite{Wang:2019urm} studied the phase
structure and transitions of Gauss-Bonnet black holes in a cavity, which are
quite similar to the AdS counterparts. However, it has been reported that
black holes in a cavity and the AdS counterparts can show different
thermodynamic behavior, e.g., phase structure of Born-Infeld black holes
\cite{Wang:2019kxp,Liang:2019dni}, the second law of thermodynamics
\cite{Wang:2020osg} and thermodynamic geometry of RN black holes
\cite{Wang:2019cax}. Furthermore, thermodynamics and critical behavior of de
Sitter black holes in a cavity were investigated in
\cite{Simovic:2018tdy,Simovic:2019zgb,Haroon:2020vpr}. To make the
thermodynamics of black holes in a cavity become more complete, we recently
extended the phase space of black holes in a cavity\ by including a
thermodynamic pressure and a thermodynamic volume \cite{Wang:2020hjw}. It
showed that, in these extended phase spaces, the thermodynamic behavior of
black holes in a cavity closely resembles that of the AdS counterparts, both
exhibiting well-known phenomena found in previous studies, such as
Hawking-Page-like and van der Waals-like phase transitions.

Although the idea that black hole can be understood as a thermodynamic system
has convinced most people, the statistical description of the black hole
microstates has not yet been fully understood, which has a deep impact upon
the understanding of quantum gravity. On the other hand, the thermodynamic
geometry method provides a useful tool to probe into microstructure of black
holes. Inspired by the pioneering work of Weinhold \cite{Weinhold1975Metric},
Ruppeiner \cite{Ruppeiner:1995zz} introduced a Riemannian thermodynamic
entropy metric to describe the thermodynamic fluctuation theory and proposed a
systematic way to calculate the Ricci curvature scalar $R$ of the Ruppeiner
metric. It showed that the sign of $R$ can relate to the type of interparticle
interactions: $R>0$ corresponds to a repulsive interaction (e.g., ideal Bose
gas), $R<0$ to an attractive interaction (e.g., ideal Fermi gas), and $R=0$ to
no interaction (e.g., ideal gas). At a critical point, $|R|$ diverges as the
correlation volume. Since the work of \cite{Ferrara:1997tw}, thermodynamic
geometry has been explored for various black holes
\cite{Aman:2003ug,Sarkar:2006tg,Quevedo:2008xn,Banerjee:2010bx,Astefanesei:2010bm,Liu:2010sz,Banerjee:2016nse,Vetsov:2018dte,Dimov:2019fxp,HosseiniMansoori:2019jcs,Bhattacharya:2019awq}%
. Interestingly, RN-AdS black holes have been proposed to be built of some
micromolecules, interactions among which can be tested by $R$
\cite{Wei:2015iwa,Wei:2019uqg}. For a RN-AdS black hole in the extended phase
space, it was observed that $R$ can be positive or negative, and the critical
behavior resembles that of ordinary thermodynamic systems
\cite{Shen:2005nu,Sahay:2010tx,Niu:2011tb,Chaturvedi:2017vgq,HosseiniMansoori:2020jrx}%
. Subsequently, thermodynamic geometry was investigated for other AdS black
holes in the extended phase space
\cite{Zhang:2015ova,Hendi:2015xya,Sahay:2016kex,Dehyadegari:2016nkd,Miao:2017cyt,Li:2017xvi,Du:2019poh,Guo:2019oad,Wei:2019yvs,Wei:2019ctz,Yerra:2020oph,Yerra:2020tzg,Yerra:2021hnh}%
.

Nevertheless, thermodynamic geometry has rarely been reported for black holes
in cavity in the context of the extended phase space. In this paper, we
investigate the thermodynamic geometry of RN black holes in cavity in the
extended phase space. The rest of this paper is organized as follows. In
section \ref{Sec:TGAdS}, we discuss the phase structure and the thermodynamic
geometry of RN-AdS black holes in the extended phase space. In section
\ref{Sec:TGCA}, the phase structure and the thermodynamic geometry of RN black
hole in a cavity are studied in the extended phase space. We summarize our
results with a brief discussion in section \ref{Sec:DC}. For simplicity, we
set $G=\hbar=c=k_{B}=1$ in this paper.

\section{RN-AdS Black Holes}

\label{Sec:TGAdS}

In this section, we study the phase structure and the thermodynamic geometry
of RN-AdS black holes in the extended phase space. The thermodynamic geometry,
also known as the Ruppeiner geometry, is an important concept in understanding
the microstructure of a thermodynamic system from its macroscopic quantities.
Adopting the Ruppeiner approach, one can define the Ruppeiner metric
$g_{\mu\nu}^{R}$ for a thermodynamic system of independent variables $x^{\mu}$
as%
\begin{equation}
g_{\mu\nu}^{R}=-\frac{\partial^{2}S\left(  x\right)  }{\partial x^{\mu
}\partial x^{\nu}}, \label{eq:RMetric}%
\end{equation}
where $S$ is the entropy of the system. Then one can define a scalar
curvature, namely the Ruppeiner invariant $R$, in this parameter space with
the Ruppeiner metric $\left(  \ref{eq:RMetric}\right)  $. Note that $R>0$
($R<0$) implies a repulsive (attractive) interaction. In this paper, we study
the Ruppeiner geometry with the internal energy and the volume being
fluctuation variables, corresponding to $x^{\mu}=\left(  U,V\right)  $.
Therefore, the Ruppeiner metric becomes \cite{Wei:2019yvs}%
\begin{equation}
g_{\mu\nu}^{R}dx^{\mu}dx^{\nu}=d\left(  \frac{1}{T}\right)  dU+d\left(
\frac{P}{T}\right)  dV=\frac{1}{T}dTdS-\frac{1}{T}dVdP. \label{eq:gTSVP}%
\end{equation}

The $4$-dimensional static charged RN-AdS black hole solution is described by%
\begin{equation}
ds^{2}=-f\left(  r\right)  dt^{2}+\frac{dr^{2}}{f\left(  r\right)  }%
+r^{2}\left(  d\theta^{2}+\sin^{2}\theta d\phi^{2}\right)  \text{, }%
A=-\frac{Q}{r}dt,
\end{equation}
where the metric function $f\left(  r\right)  $ is
\begin{equation}
f\left(  r\right)  =1-\frac{2M}{r}+\frac{Q^{2}}{r^{2}}+\frac{r^{2}}{l^{2}},
\label{eq:frAdS}%
\end{equation}
and $l$ is the AdS radius. Here, the parameters $M$ and $Q$ are the black hole
mass and charge, respectively. The Hawking temperature $T$ is given by%
\begin{equation}
T=\frac{1}{4\pi r_{+}}\left(  1+\frac{3r_{+}^{2}}{l^{2}}-\frac{Q^{2}}%
{r_{+}^{2}}\right)  ,
\end{equation}
where $r_{+}$ is the radius of the event horizon, satisfying $f(r_{+})=0$. The
mass $M$ can also be expressed in terms of $r_{+}$,%
\begin{equation}
M=\frac{r_{+}}{2}\left(  1+\frac{Q^{2}}{r_{+}^{2}}+\frac{r_{+}^{2}}{l^{2}%
}\right)  .
\end{equation}
Defining the pressure \cite{Dolan:2011xt},%
\begin{equation}
P\equiv\frac{3}{8\pi l^{2}}, \label{eq:P}%
\end{equation}
one can have%
\begin{equation}
T=\frac{1}{4\pi r_{+}}\left(  1+8\pi r_{+}^{2}P-\frac{Q^{2}}{r_{+}^{2}%
}\right)  ,\text{ }M=\frac{r_{+}}{2}\left(  1+\frac{Q^{2}}{r_{+}^{2}}+\frac
{8}{3}\pi r_{+}^{2}P\right)  . \label{eq:TM}%
\end{equation}
Moreover, the entropy obeys the Bekenstein-Hawking area entropy relation,%
\begin{equation}
S=\frac{A}{4}=\pi r_{+}^{2},
\end{equation}
and the thermodynamic volume is given by%
\begin{equation}
V\equiv\frac{\partial M}{\partial P}=\frac{4}{3}\pi r_{+}^{3}. \label{eq:V}%
\end{equation}
For later convenience, we rescale thermodynamic variables,
\begin{equation}
\tilde{r}_{+}\equiv r_{+}/Q,\text{ }\tilde{T}\equiv TQ,\text{ }\tilde{P}\equiv
PQ^{2},\text{ }\tilde{V}\equiv V/Q^{3},\text{ }\tilde{G}\equiv G/Q,
\label{eq:scale}%
\end{equation}
where $G=M-TS$ is the Gibbs free energy, and tilde quantities are
dimensionless. The rescaled thermodynamic variables can be expressed as%
\begin{equation}
\tilde{V}\left(  \tilde{r}_{+}\right)  =\frac{4}{3}\pi\tilde{r}_{+}^{3}%
,\tilde{T}(\tilde{r}_{+},\tilde{P})=\frac{1}{4\pi\tilde{r}_{+}}\left(
1+8\pi\tilde{r}_{+}^{2}\tilde{P}-\frac{1}{\tilde{r}_{+}^{2}}\right)
,\tilde{G}(\tilde{r}_{+},\tilde{P})=\frac{1}{12\tilde{r}_{+}}\left(
9+3\tilde{r}_{+}^{2}-8\pi\tilde{r}_{+}^{4}\tilde{P}\right)  . \label{eq:VTGQ}%
\end{equation}

\begin{figure}[tb]
\begin{center}
\includegraphics[width=0.4\textwidth]{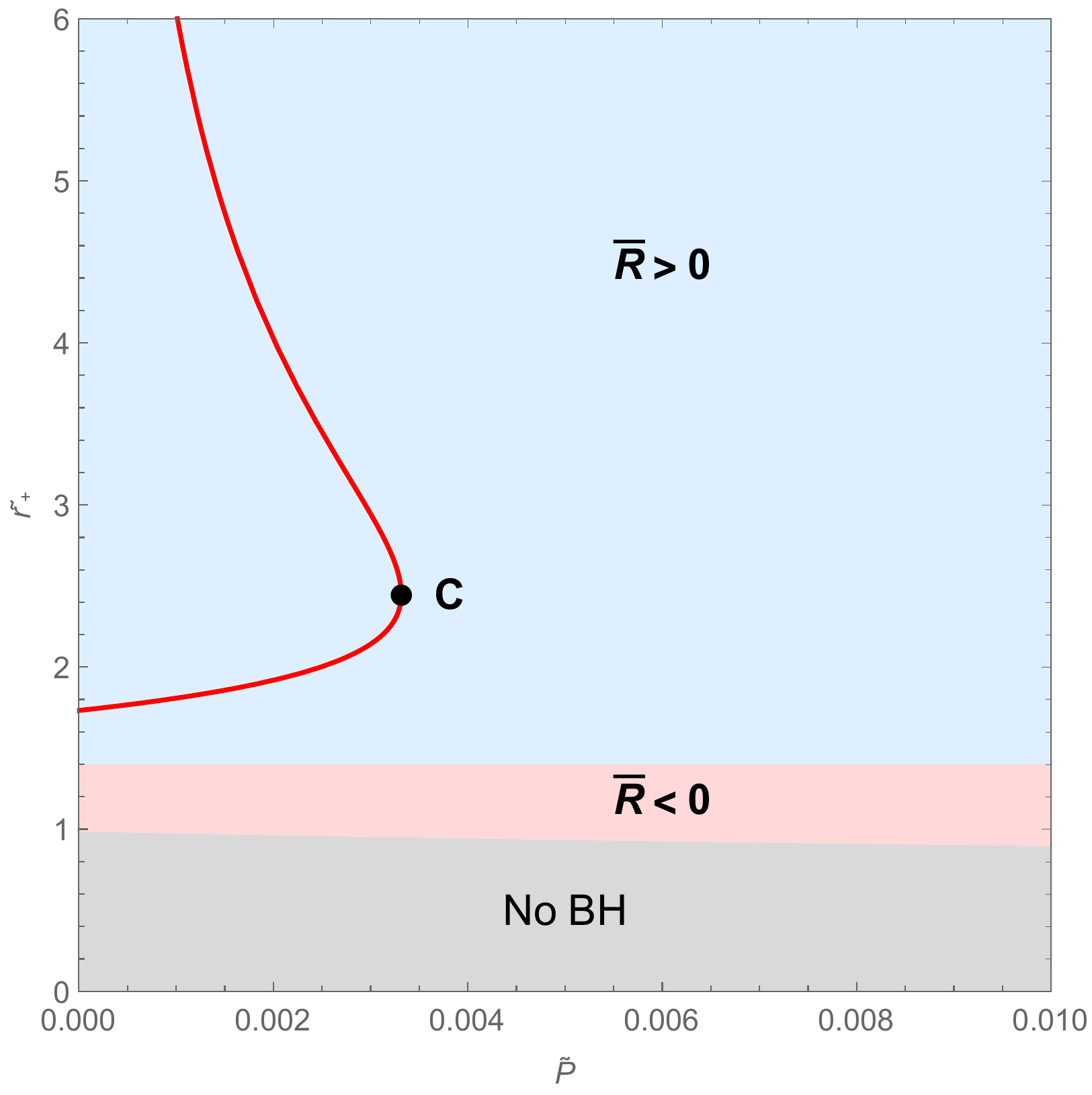}
\end{center}
\caption{The region-plot of the Ruppeiner invariant for RN-AdS black holes in
the $\tilde{P}$-$\tilde{r}_{+}$ plane, where $\tilde{r}_{+}$ is the rescaled
event horizon radius and $\tilde{P}$ is the rescaled pressure. In the red
(blue) region, $\bar{R}<0$ $(\bar{R}>0)$. The red line corresponds to $\bar
{R}=+\infty.$ The black dot $C$ denotes the critical point, which lies on the
$\bar{R}=+\infty$ line. Black hole solutions do not exist in the gray region.}%
\label{fig:rPR0AdS}%
\end{figure}

For the convenience of calculation, one can rewrite the Ruppeiner metric
$\left(  \ref{eq:gTSVP}\right)  $ in the $(T,V)$ coordinate \cite{Wei:2019yvs}%
,
\begin{equation}
g_{\mu\nu}^{R}dx^{\mu}dx^{\nu}=\frac{1}{T^{2}}\left[  C_{V}dT^{2}-T\left(
\frac{\partial P}{\partial V}\right)  _{T}dV^{2}\right]  . \label{eq:gTV}%
\end{equation}
However, $C_{V}=T\left(  \frac{\partial S}{\partial T}\right)  _{V}=0$ for a
RN-AdS black hole, which gives that the metric is not invertible, and the
Ruppeiner invariant $R$ diverges. Compared with the specific heat capacity of
the Van der Waals (VdW) fluid, $C_{V}^{VdW}=3k_{B}/2$, the specific heat
capacity of the RN-AdS black hole can be treated as the limit of the VdW fluid
with $k_{B}\rightarrow0$. Then one can regularize the divergent Ruppeiner
invariant by normalizing it with $C_{V}$ \cite{Wei:2019yvs}. In other words,
one can treat $C_{V}$ as a constant whose value is infinitesimally close to
zero and define a normalized Ruppeiner invariant,%
\begin{equation}
\bar{R}(\tilde{r}_{+},\tilde{P})\equiv RC_{V}=\frac{2\left(  \tilde{r}_{+}%
^{2}-2\right)  \left(  8\pi\tilde{r}_{+}^{4}\tilde{P}+1\right)  }{\left(
\tilde{r}_{+}^{2}-3-8\pi\tilde{r}_{+}^{4}\tilde{P}\right)  ^{2}}.
\label{eq:RTVC}%
\end{equation}
In FIG. \ref{fig:rPR0AdS}, we present the region-plot of $\bar{R}$ in the
$\tilde{P}$-$\tilde{r}_{+}$ plane, where $\bar{R}<0$ $(\bar{R}>0)$ in the red
(blue) region, and $\bar{R}\rightarrow+\infty$ on the red line. The critical
point, denoted by a black dot, lies on the $\bar{R}=+\infty$ line. Note that
black hole solutions do not exist in the gray region, where $T$ is negative.
And the interaction is attractive for small enough horizon radius, while the
interaction is repulsive for large enough horizon radius. 

To study phase structure, we consider the Gibbs free energy
\begin{equation}
\tilde{G}(\tilde{r}_{+},\tilde{P})=\frac{1}{12\tilde{r}_{+}}\left(
9+3\tilde{r}_{+}^{2}-8\tilde{P}\pi\tilde{r}_{+}^{4}\right)  . \label{eq:GQrP}%
\end{equation}
Solving eqn. $\left(  \ref{eq:VTGQ}\right)  $ for $\tilde{r}_{+}$ in terms of
$\tilde{T}$ gives $\tilde{r}_{+}=\tilde{r}_{+}(\tilde{T},\tilde{P})$. It shows
that $\tilde{r}_{+}(\tilde{T},\tilde{P})$ is multi-valued for small enough
$\tilde{P}$ and single-valued for large enough $\tilde{P}$, which indicates
that there is a critical point. The critical point is determined by
\begin{equation}
\frac{\partial\tilde{T}(\tilde{r}_{+},\tilde{P})}{\partial\tilde{r}_{+}%
}=0\text{, }\frac{\partial^{2}\tilde{T}(\tilde{r}_{+},\tilde{P})}%
{\partial\tilde{r}_{+}^{2}}=0,
\end{equation}
which gives the critical values,
\begin{equation}
\tilde{r}_{+c}=\sqrt{6}\text{, }\tilde{P}_{c}=\frac{1}{96\pi}\text{, }%
\tilde{T}_{c}=\frac{1}{3\sqrt{6}\pi}.
\end{equation}
If $\tilde{r}_{+}(\tilde{T},\tilde{P})$ is multi-valued, there is more than
one black hole solution for fixed values of $\tilde{P}$ and $\tilde{T}$,
corresponding to multiple phases in a canonical ensemble. Plugging $\tilde
{r}_{+}(\tilde{T},\tilde{P})$ into eqns. $\left(  \ref{eq:RTVC}\right)  $ and
$\left(  \ref{eq:GQrP}\right)  $, one can have $\tilde{G}(\tilde{T},\tilde
{P})$ and $\bar{R}(\tilde{T},\tilde{P})$. \begin{figure}[tb]
\begin{center}
\includegraphics[width=0.4\textwidth]{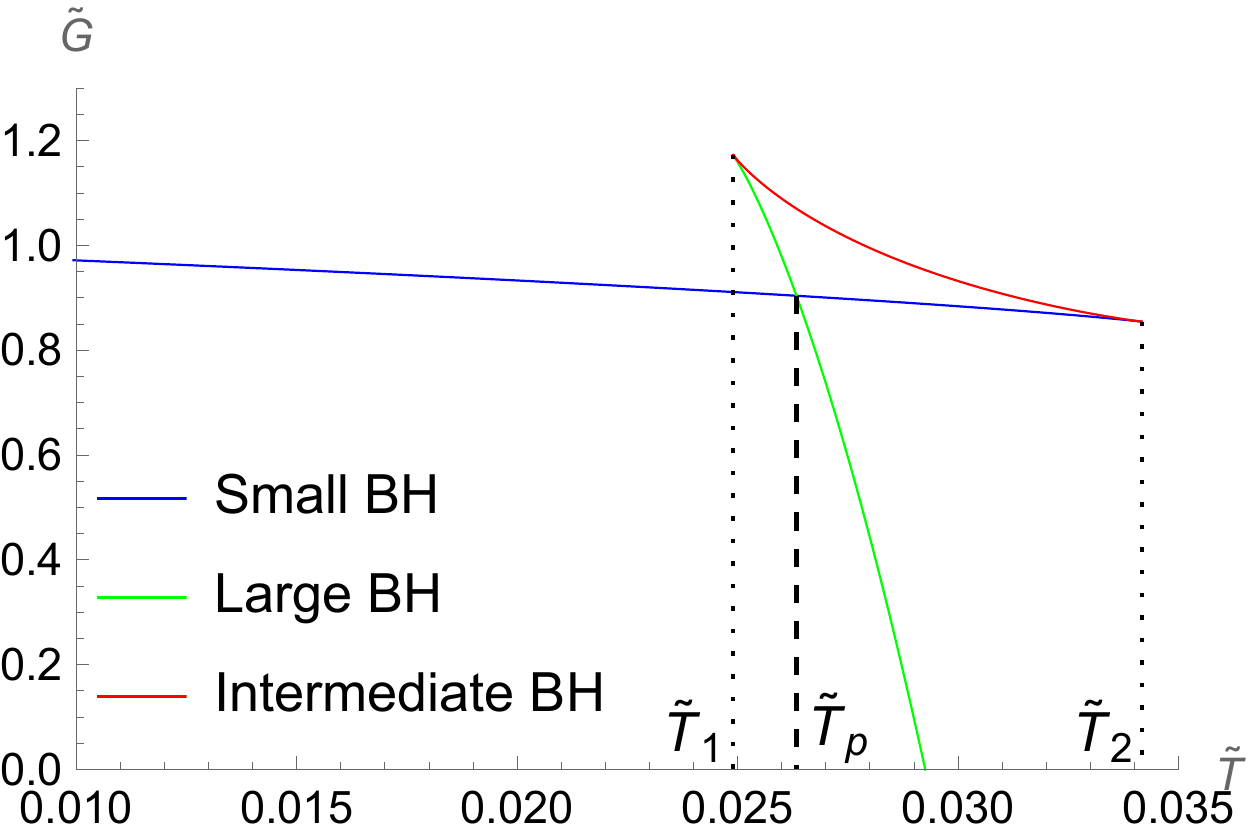}\label{fig:GRTAdS:a}
\hspace{1cm}
\includegraphics[width=0.4\textwidth]{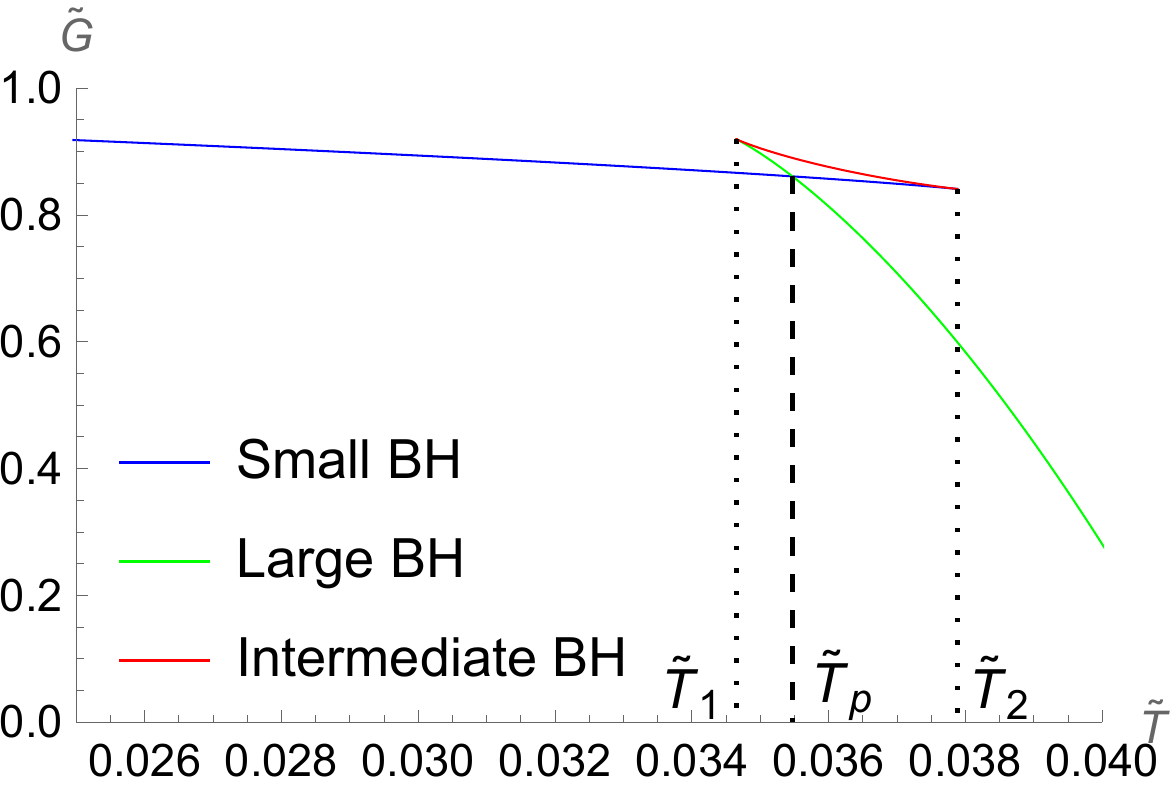}\label{fig:GRTAdS:b}
\includegraphics[width=0.4\textwidth]{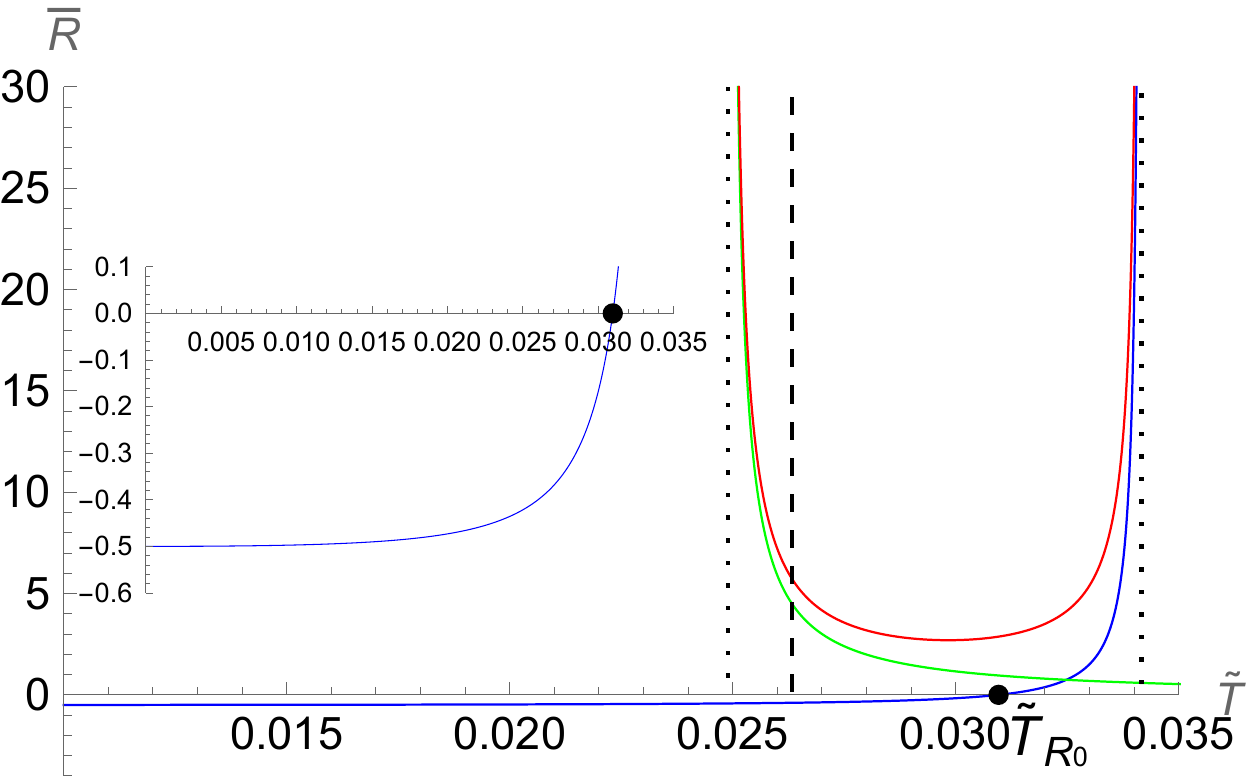}\label{fig:GRTAdS:c}
\hspace{1cm}
\includegraphics[width=0.4\textwidth]{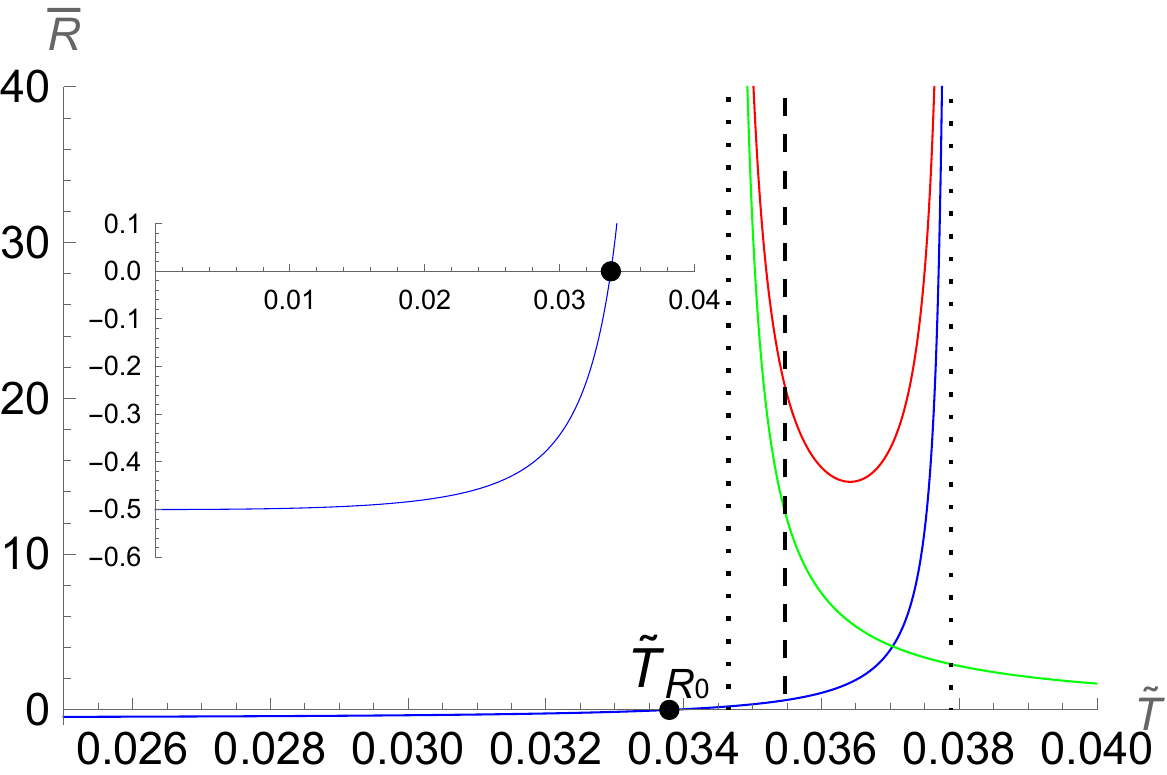}\label{fig:GRTAdS:d}
\end{center}
\caption{Plots of the rescaled Gibbs free energy $\tilde{G}\ $and the
normalized Ruppeiner invariant $\bar{R}$\ against the rescaled temperature
$\tilde{T}$\ for RN-AdS black holes when $\tilde{P}<\tilde{P}_{c}$. There are
three black hole solutions when $\tilde{T}_{1}<\tilde{T}<\tilde{T}_{2}$. As
$\tilde{T}$ increases from $0$, a first-order transition from Small BH to
Large BH occurs at $\tilde{T}=\tilde{T}_{p}$. The black points correspond to
$\bar{R}=0$ at $\tilde{T}=\tilde{T}_{R_{0}}$. \textbf{Left Column}: $\tilde
{P}=0.001$, one has $\tilde{T}_{R_{0}}>\tilde{T}_{p}$. As one increases
$\tilde{T}$ from $0$, the attractive interaction turns repulsive when the
phase transition occurs, and $\bar{R}$ cannot be $0$ for the thermodynamically
preferred phase. \textbf{Right Column}: $\tilde{P}=0.002$, one has $\tilde
{T}_{R_{0}}<\tilde{T}_{p}$. As one increases $\tilde{T}$ from $0$, the
attractive interaction turns repulsive before the phase transition occurs.}%
\label{fig:GRTAdSS}%
\end{figure}

When $\tilde{P}<\tilde{P}_{c}$, there are three solutions coexisting for some
range of $\tilde{T}$. In fact, we plot $\tilde{G}$ and $\bar{R}$ against
$\tilde{T}$ in FIG. \ref{fig:GRTAdSS}, which shows that three solutions,
dubbed Large BH, Small BH and Intermediate BH, coexist for $\tilde{T}%
_{1}<\tilde{T}<\tilde{T}_{2}$. Intermediate BH is a thermally unstable phase,
while Small and Large BHs are thermally stable. The upper row of FIG.
\ref{fig:GRTAdSS} shows that there is a first-order phase transition between
Small BH and Large BH at $\tilde{T}=\tilde{T}_{p}$ with $\tilde{T}_{1}%
<\tilde{T}_{p}<\tilde{T}_{2}.$ The bottom row displays that $\bar{R}$ of Large
BH and Intermediate BH are always positive, corresponding to repulsive
interactions between BH molecules. For Large BH and Small BH, $\bar
{R}\rightarrow\infty$ at $\tilde{T}=\tilde{T}_{1}$ and $\tilde{T}=\tilde
{T}_{2}$, respectively. Moreover, for Small BH, $\bar{R}=-1/2$ for $\tilde
{T}=0$, and $\bar{R}\rightarrow\infty$ for $\tilde{T}=\tilde{T}_{2}$,
indicating that there exists $\bar{R}=0$ at $\tilde{T}=\tilde{T}_{R_{0}}$.
Furthermore, we find that $\tilde{T}_{R_{0}}<\tilde{T}_{p}$ when $\tilde
{P}>\tilde{P}_{I}=3/[32(9+4\sqrt{5})\pi]$, and $\tilde{T}_{R_{0}}>\tilde
{T}_{p}$ when $\tilde{P}<\tilde{P}_{I}$. The case with $\tilde{P}%
=0.001<\tilde{P}_{I}$ is shown in the left column of FIG. \ref{fig:GRTAdSS},
which displays that the phase transition occurs before $\tilde{T}=\tilde
{T}_{R_{0}}$ as $\tilde{T}$ increases from $0$. Since $\bar{R}$ is always
positive for Large BH, the attractive interactions between the BH molecules
turns into the repulsive one when the phase transition occurs. For $\tilde
{P}_{I}<\tilde{P}<\tilde{P}_{c}$, the right column of FIG. \ref{fig:GRTAdSS}
shows that as one increases $\tilde{T}$ from $0$, the attractive interaction
turns into the repulsive one before the phase transition occurs.
\begin{figure}[tb]
\begin{center}
\includegraphics[width=0.4\textwidth]{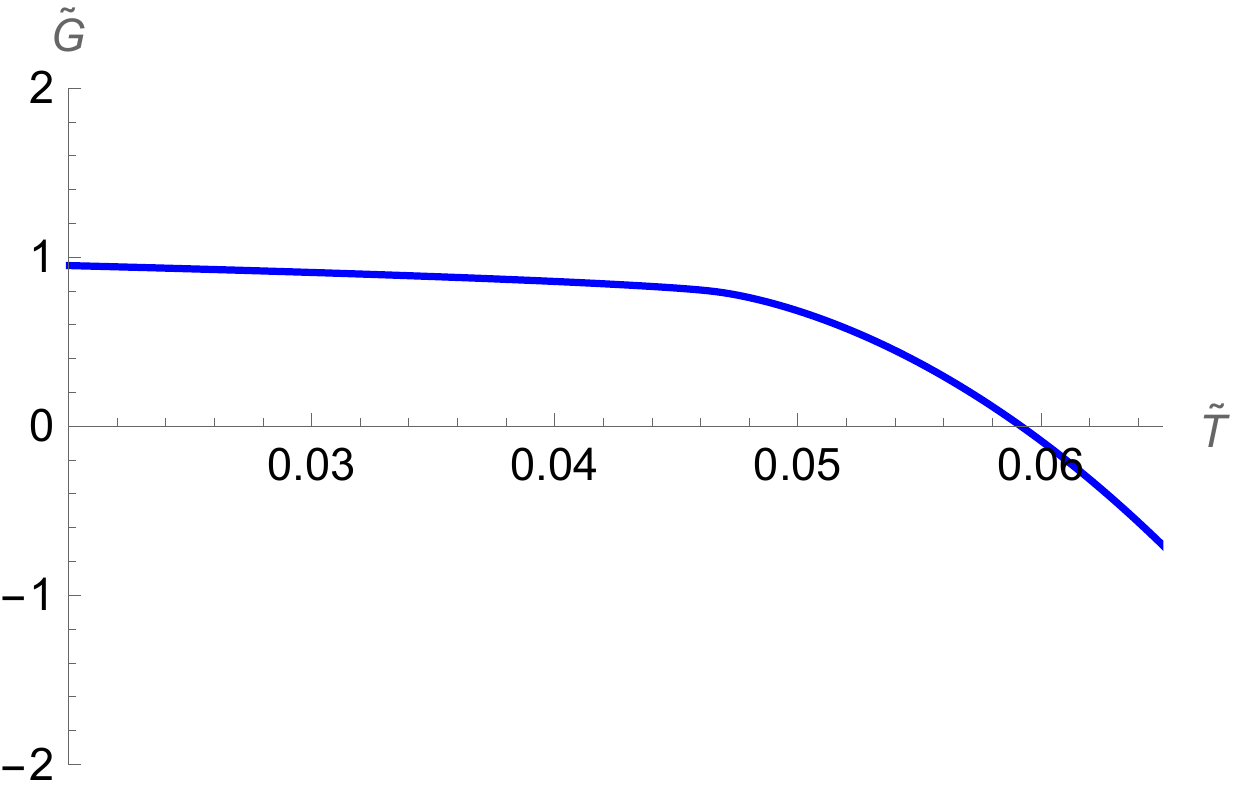}\label{fig:GRTAdSL:a}
\hspace{1cm}
\includegraphics[width=0.4\textwidth]{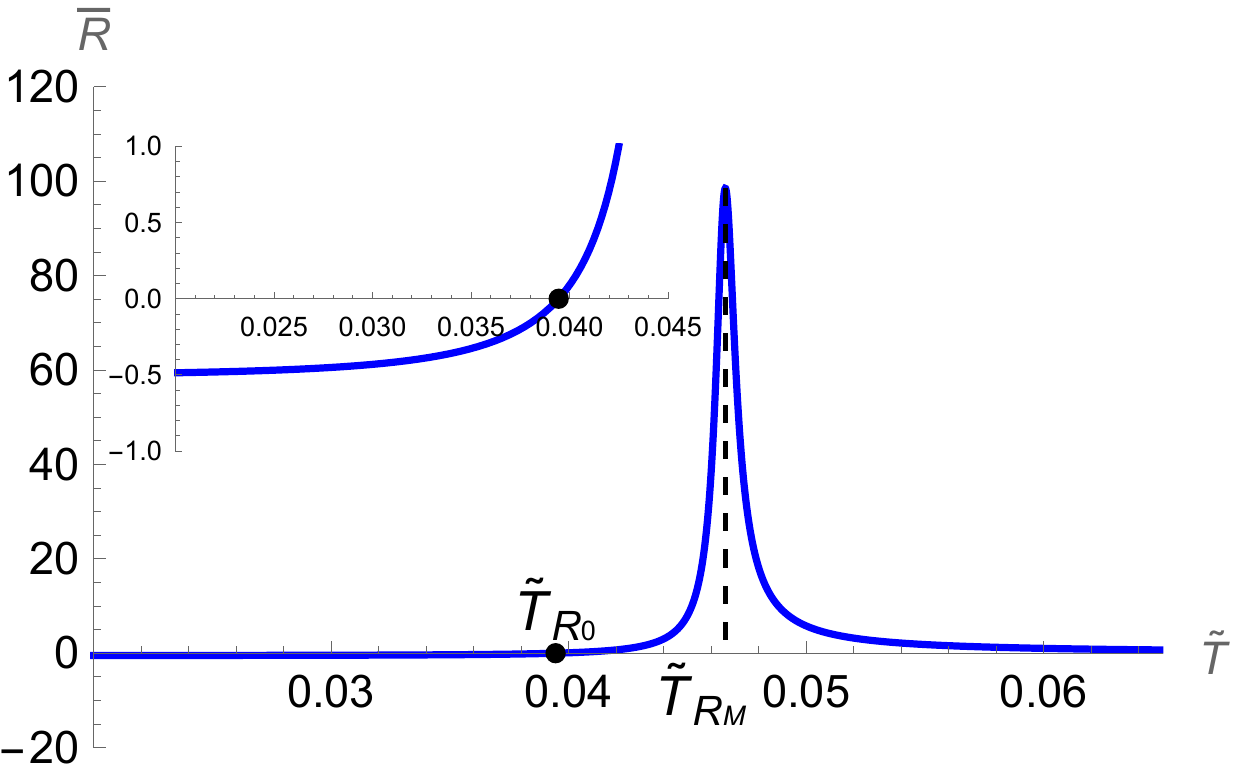}\label{fig:GRTAdSL:b}
\end{center}
\caption{Plots of the rescaled Gibbs free energy $\tilde{G}\ $and the
normalized Ruppeiner invariant $\bar{R}$\ against the rescaled temperature
$\tilde{T}$\ for RN-AdS black holes with $\tilde{P}=0.004>\tilde{P}_{c}$.
There is only one branch and no phase transition. The normalized Ruppeiner
invariant $\bar{R}=-1/2$ when $\tilde{T}=0,$ and $\bar{R}\rightarrow0$ when
$\tilde{T}\rightarrow\infty$. The black point corresponds to $\bar{R}=0$. And
$\bar{R}$ has a maximum at $\tilde{T}=\tilde{T}_{R_{M}}$.}%
\label{fig:GRTAdSL}%
\end{figure}

When $\tilde{P}>\tilde{P}_{c}$, there is a single thermally stable solution
for a fixed $\tilde{T}$. We plot $\tilde{G}$ and $\bar{R}$ against $\tilde{T}$
for $\tilde{P}=0.004$ in FIG. \ref{fig:GRTAdSL}. The right panel of FIG.
\ref{fig:GRTAdSL} displays that $\bar{R}=0$ at $\tilde{T}=\tilde{T}_{R_{0}}$,
and $\bar{R}$ has a maximum at $\tilde{T}=\tilde{T}_{R_{M}}$. As one increases
$\tilde{T}$ from $0$, $\bar{R}$ becomes $0$ at $\tilde{T}=\tilde{T}_{R_{0}}<$
$\tilde{T}_{R_{M}}$, and $\bar{R}$ is always positive for $\tilde{T}>$
$\tilde{T}_{R_{0}}$, which means that the attractive interaction turns
repulsive at $\tilde{T}=\tilde{T}_{R_{0}}$.

\begin{figure}[tb]
\begin{center}
\includegraphics[width=0.4\textwidth]{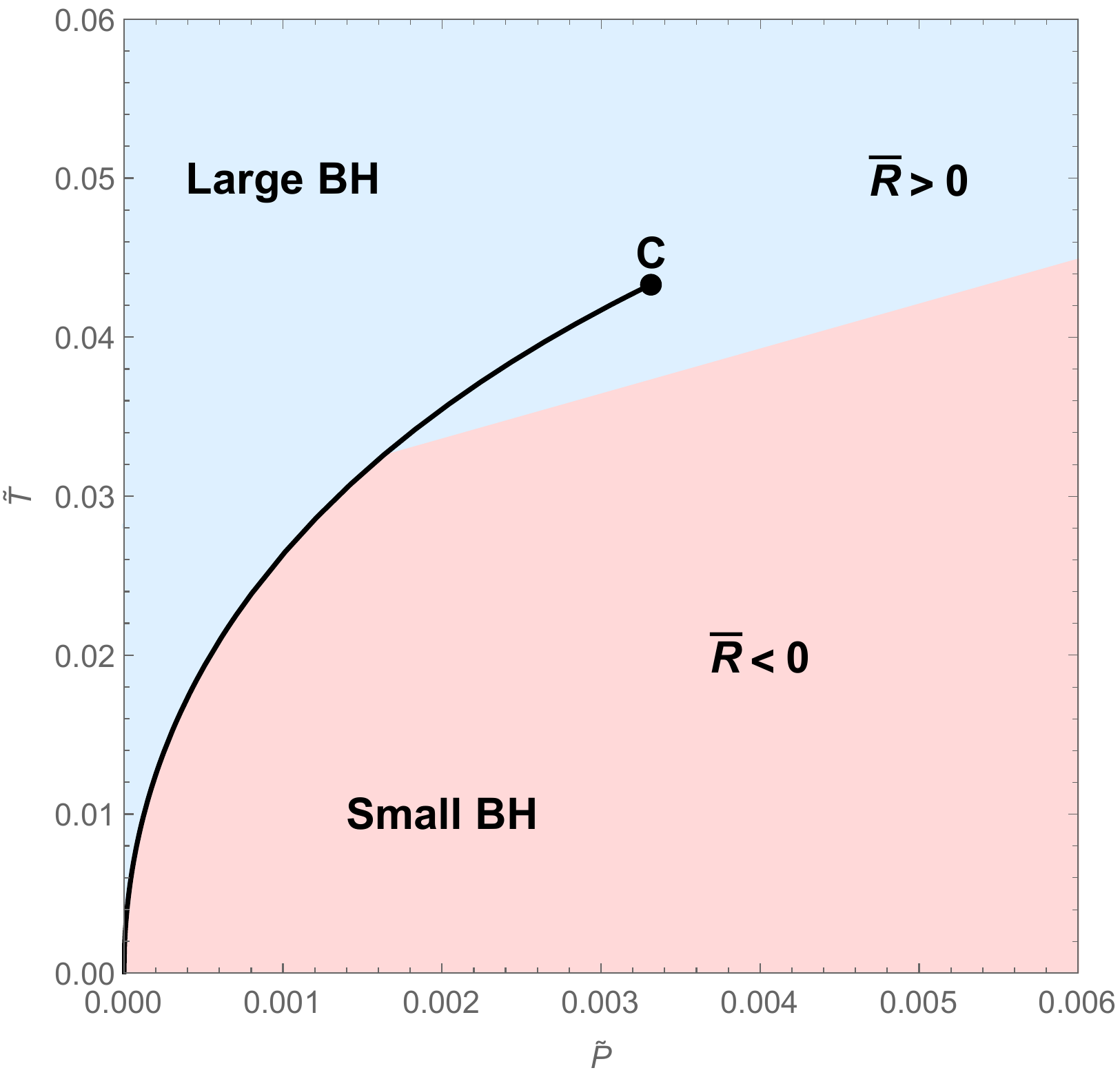}
\end{center}
\caption{Phase diagram of RN-AdS black holes in the $\tilde{P}$-$\tilde{T}$
plane. The first-order phase transition line, separating Large BH and Small
BH, is displayed by a black line and terminates at the critical point, marked
by a black dot. The blue region and red region correspond to repulsive and
attractive interactions, respectively.}%
\label{fig:TPR0AdS}%
\end{figure}

In FIG. \ref{fig:TPR0AdS}, we display the globally preferred phases of RN-AdS
black holes, and the regions of $\bar{R}>0$ and $\bar{R}<0$ in the $\tilde{P}%
$-$\tilde{T}$ plane. The black line is the first-order phase transition line,
separating Large BH and Small BH. And it terminates at a critical point,
marked by a black dot. In the blue (red) region, one has $\bar{R}>0$ $\left(
\bar{R}<0\right)  $, which means that the type of interactions between BH
molecules is repulsive (attractive). Note that $\bar{R}$ is always finite in
the $\tilde{P}$-$\tilde{T}$ plane.

\section{RN Black Holes in a Cavity}

\label{Sec:TGCA}

In this section, we study the phase structure and the thermodynamic geometry
of RN black holes enclosed by a spherical cavity in an extended phase space.
The $4$-dimensional RN black hole solution is%
\begin{equation}
ds^{2}=-f\left(  r\right)  dt^{2}+\frac{dr^{2}}{f\left(  r\right)  }%
+r^{2}\left(  d\theta^{2}+\sin^{2}\theta d\phi^{2}\right)  ,
\end{equation}
where the metric function is
\begin{equation}
f\left(  r\right)  =\left(  1-\frac{r_{+}}{r}\right)  \left(  1-\frac
{Q_{b}^{2}}{r_{+}r}\right)  ,
\end{equation}
and the electromagnetic potential is%
\begin{equation}
A=-\frac{Q_{b}}{r}dt.
\end{equation}
Here, $Q_{b}$ is the black hole charge, and $r_{+}$ is the radius of the event
horizon. The Hawking temperature $T_{b}$ of the RN black hole is given by%
\begin{equation}
T_{b}=\frac{1}{4\pi r_{+}}\left(  1-\frac{Q_{b}^{2}}{r_{+}^{2}}\right)  .
\end{equation}
The wall of the cavity, which is located at $r=r_{B}$, is maintained at a
temperature of $T$ and a charge of $Q$. It can show that the system
temperature $T$ and charge $Q$ can be related to the black hole temperature
$T_{b}$ and charge $Q_{b}$ as \cite{Braden:1990hw}%
\begin{equation}
Q=Q_{b}\text{ and }T=\frac{T_{b}}{\sqrt{f\left(  r_{B}\right)  }},
\label{eq:RNTQ}%
\end{equation}
respectively. Moreover, the Helmholtz free energy $F$ and the thermal energy
$E$ were also given in \cite{Carlip:2003ne},
\begin{equation}
F=r_{B}\left[  1-\sqrt{f\left(  r_{B}\right)  }\right]  -\pi Tr_{+}^{2},\text{
}E=r_{B}\left[  1-\sqrt{f\left(  r_{B}\right)  }\right]  .
\end{equation}

In \cite{Wang:2020hjw}, we introduced a thermodynamic volume,
\begin{equation}
V\equiv4\pi r_{B}^{3}/3, \label{eq:CavityV}%
\end{equation}
and the conjugate thermodynamic pressure,%
\begin{equation}
P=-\partial E/\partial V,
\end{equation}
as a pair of extra thermodynamic variables for black holes in a cavity. In
this extended phase space, the Gibbs free energy, $G=F+PV$, is considered for
a constant pressure system. Similar to the AdS case, we also consider the
Ruppeiner metric $\left(  \ref{eq:gTSVP}\right)  $ and use eqns. $\left(
\ref{eq:RNTQ}\right)  $ and $\left(  \ref{eq:CavityV}\right)  $ to calculate
the Ruppeiner invariant $R$. Unlike the RN-AdS black holes, we find that the
Ruppeiner invariant $R$ of RN black holes in a cavity does not need any regularization.

\begin{figure}[t]
\begin{center}
\includegraphics[width=0.4\textwidth]{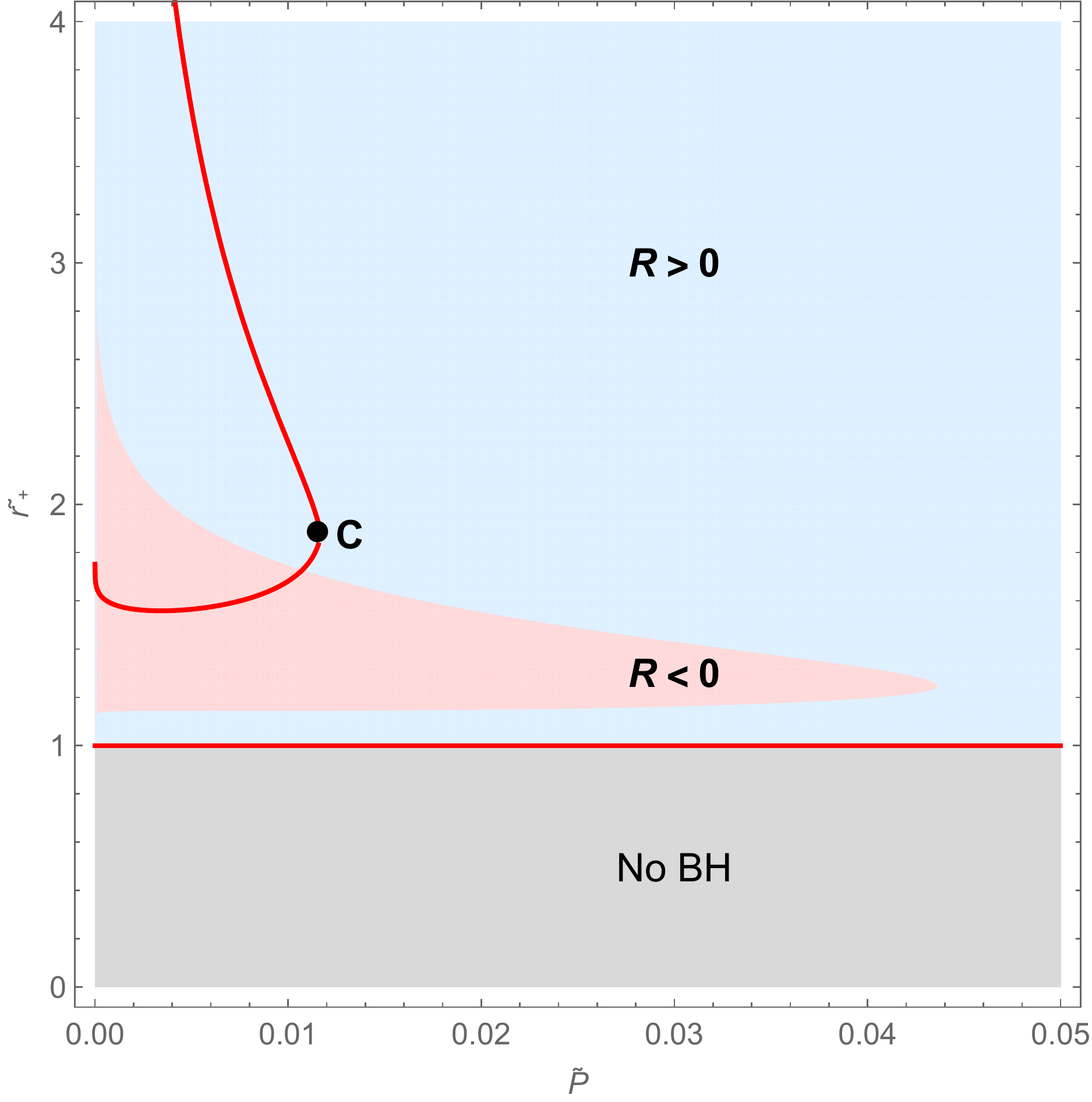}
\end{center}
\caption{The region-plot of the Ruppeiner invariant $R$ for RN black holes in
a cavity in the $\tilde{P}$-$\tilde{r}_{+}$ plane, where $\tilde{r}_{+}$ is
the rescaled event horizon radius, and $\tilde{P}$ is the rescaled pressure.
In the red (blue) region, one has $R<0$ $(R>0)$. The red line in the red
region correspond to $R\rightarrow-\infty$, and that in the blue region to
$R\rightarrow+\infty$. The critical point is marked by $C$ and lies on the
$R\rightarrow+\infty$ line. Black hole solutions do not exist in the gray
region since RN black holes in a cavity have a minimum temperature for a given
$\tilde{P}.$}%
\label{fig:rPR0Cav}%
\end{figure}

Defining rescaled thermodynamic variables,
\begin{equation}
\tilde{r}_{+}\equiv r_{+}/Q\text{, }\tilde{r}_{B}\equiv r_{B}/Q\text{, }%
\tilde{T}\equiv TQ\text{, }\tilde{P}\equiv PQ^{2}\text{, }\tilde{V}\equiv
V/Q^{3}\text{, }\tilde{G}\equiv G/Q\text{, }\tilde{R}=RQ^{2}\text{,}%
\end{equation}
we obtain
\begin{align}
\tilde{T}\left(  \tilde{r}_{+},\tilde{r}_{B}\right)   &  =\frac{1}{4\pi
\tilde{r}_{+}}\left(  1-\frac{1}{\tilde{r}_{+}^{2}}\right)  \sqrt{\frac
{\tilde{r}_{B}^{2}\tilde{r}_{+}}{\left(  \tilde{r}_{B}-\tilde{r}_{+}\right)
\left(  \tilde{r}_{+}\tilde{r}_{B}-1\right)  }},\nonumber\\
\tilde{P}\left(  \tilde{r}_{+},\tilde{r}_{B}\right)   &  =\frac{-1+2\tilde
{r}_{+}\tilde{r}_{B}-\tilde{r}_{+}^{2}-2\sqrt{\tilde{r}_{+}\left(  \tilde
{r}_{B}-\tilde{r}_{+}\right)  \left(  \tilde{r}_{+}\tilde{r}_{B}-1\right)  }%
}{8\pi\tilde{r}_{B}^{2}\sqrt{\tilde{r}_{+}\left(  \tilde{r}_{B}-\tilde{r}%
_{+}\right)  \left(  \tilde{r}_{+}\tilde{r}_{B}-1\right)  }}.
\label{eq:CavityTP}%
\end{align}
Solving $\tilde{P}\left(  \tilde{r}_{+},\tilde{r}_{B}\right)  =\tilde{P}$ for
$\tilde{r}_{B}$ in terms of $\tilde{r}_{+}$ and $\tilde{P}$ gives $\tilde
{r}_{B}=\tilde{r}_{B}(\tilde{r}_{+},\tilde{P})$, via which we can express the
rescaled Ruppeiner invariant $\tilde{R}$ as a function of $\tilde{r}_{+}$ and
$\tilde{P}$. In FIG. \ref{fig:rPR0Cav}, we display the region-plot of
$\tilde{R}$ in the $\tilde{P}$-$\tilde{r}_{+}$ plane. In the red and blue
regions, one has $\tilde{R}<0$ and $\tilde{R}>0$, respectively. Moreover,
$\tilde{R}=\pm\infty$ corresponds to the red lines. Note that $\tilde
{R}=+\infty$ at the critical point which is marked by a black point and
determined by%
\begin{equation}
\frac{\partial\tilde{T}(\tilde{r}_{+},\tilde{P})}{\partial\tilde{r}_{+}%
}=0\text{, }\frac{\partial^{2}\tilde{T}(\tilde{r}_{+},\tilde{P})}%
{\partial\tilde{r}_{+}^{2}}=0.
\end{equation}
The associated critical values are obtained by solving the above equations
\begin{equation}
\tilde{r}_{+c}=1.888\text{, }\tilde{P}_{c}=0.0115\text{, }\tilde{T}_{c}=0.118.
\end{equation}
Black hole solutions do not exist in the gray region since a RN black hole in
a cavity has a minimum of temperature $T_{\min}$ for a nonzero $\tilde{P}$
\cite{Wang:2020hjw}. Unlike RN-AdS black holes, the Ruppeiner invariant $R$ of
RN black holes in a cavity can be negative infinity in some parametric
region.\begin{figure}[tb]
\begin{center}
\includegraphics[width=0.4\textwidth]{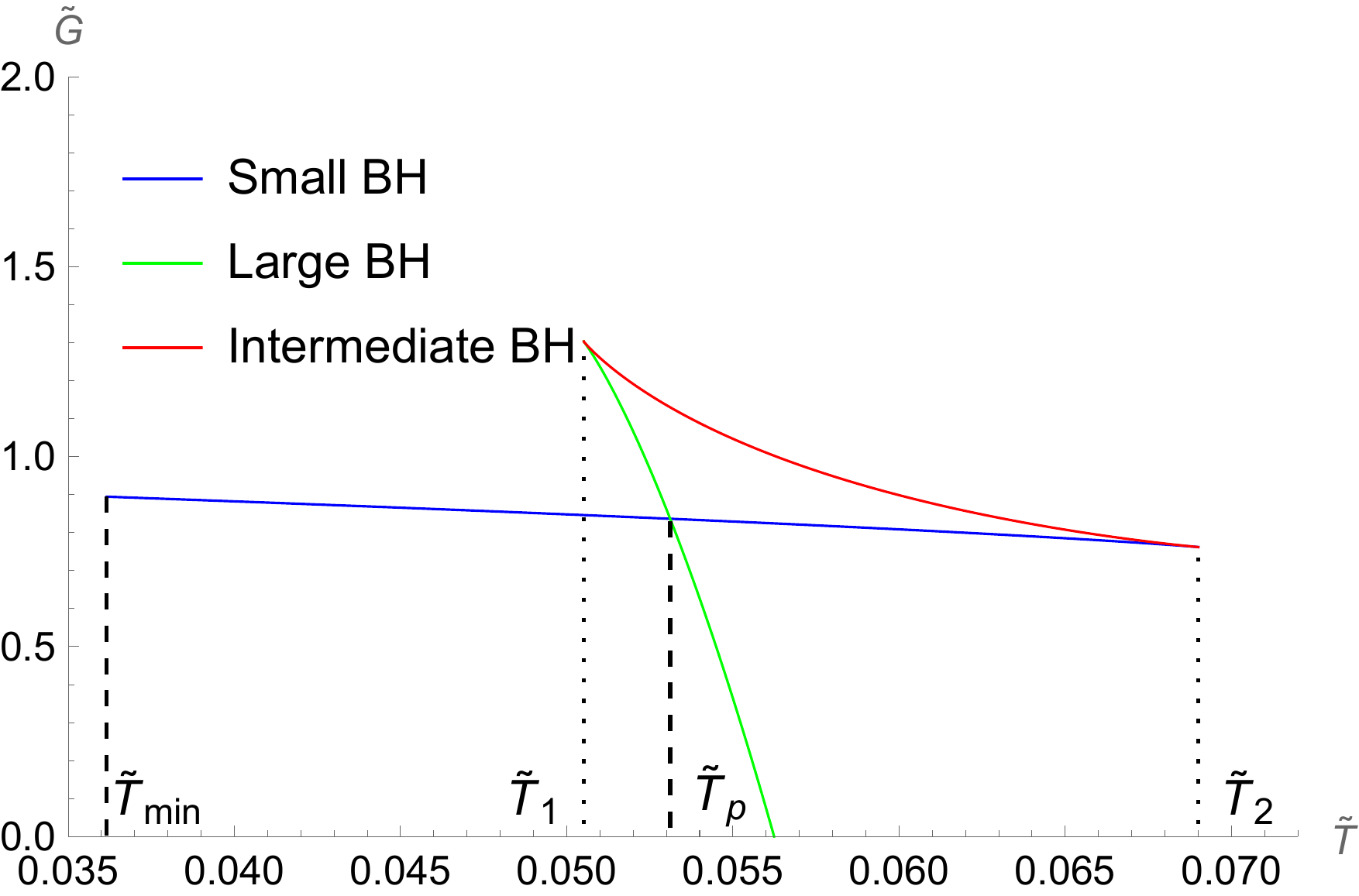}\label{fig:GRTCavS1:a}
\hspace{1cm}
\includegraphics[width=0.4\textwidth]{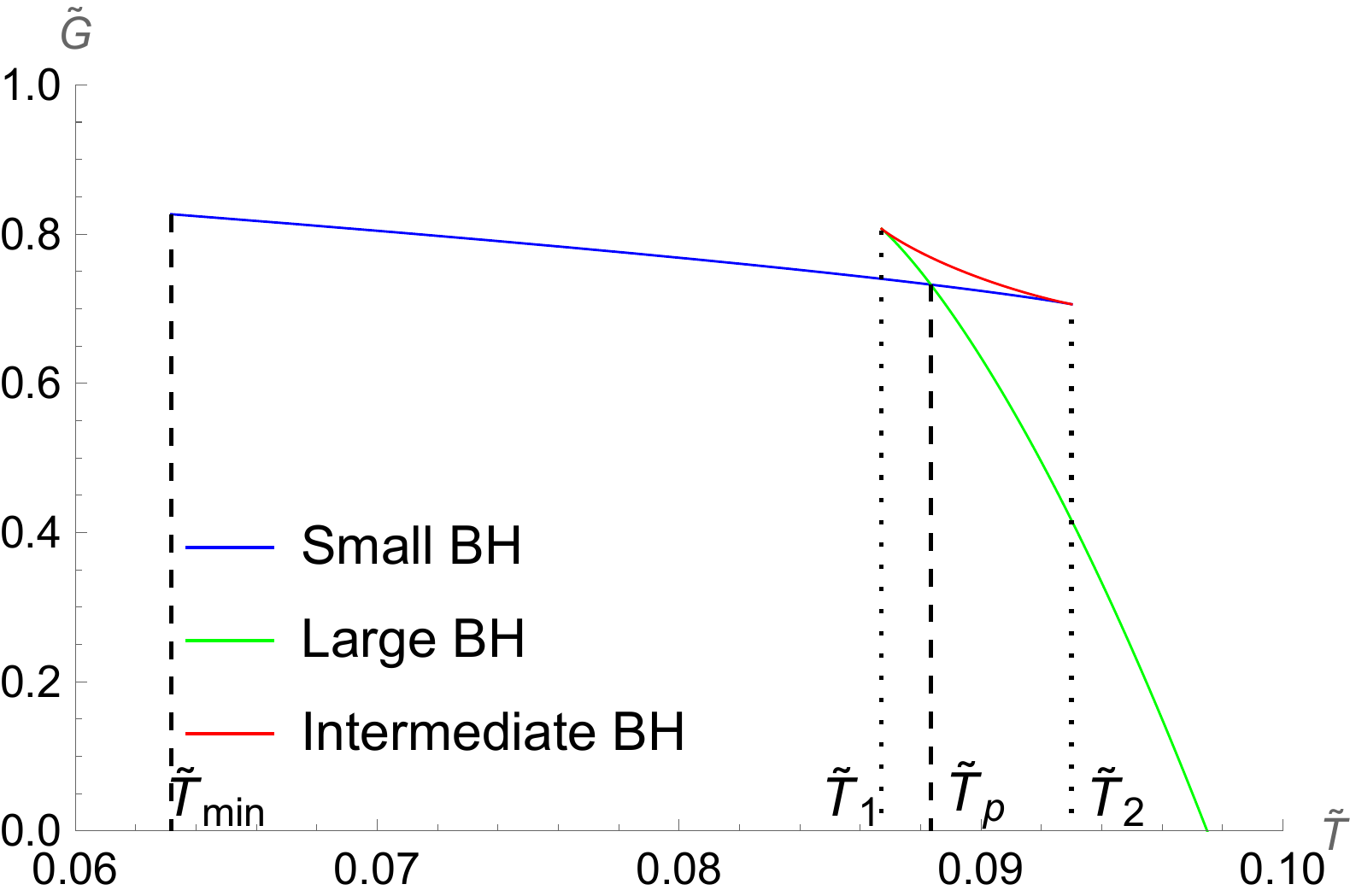}\label{fig:GRTCavS1:b}
\includegraphics[width=0.4\textwidth]{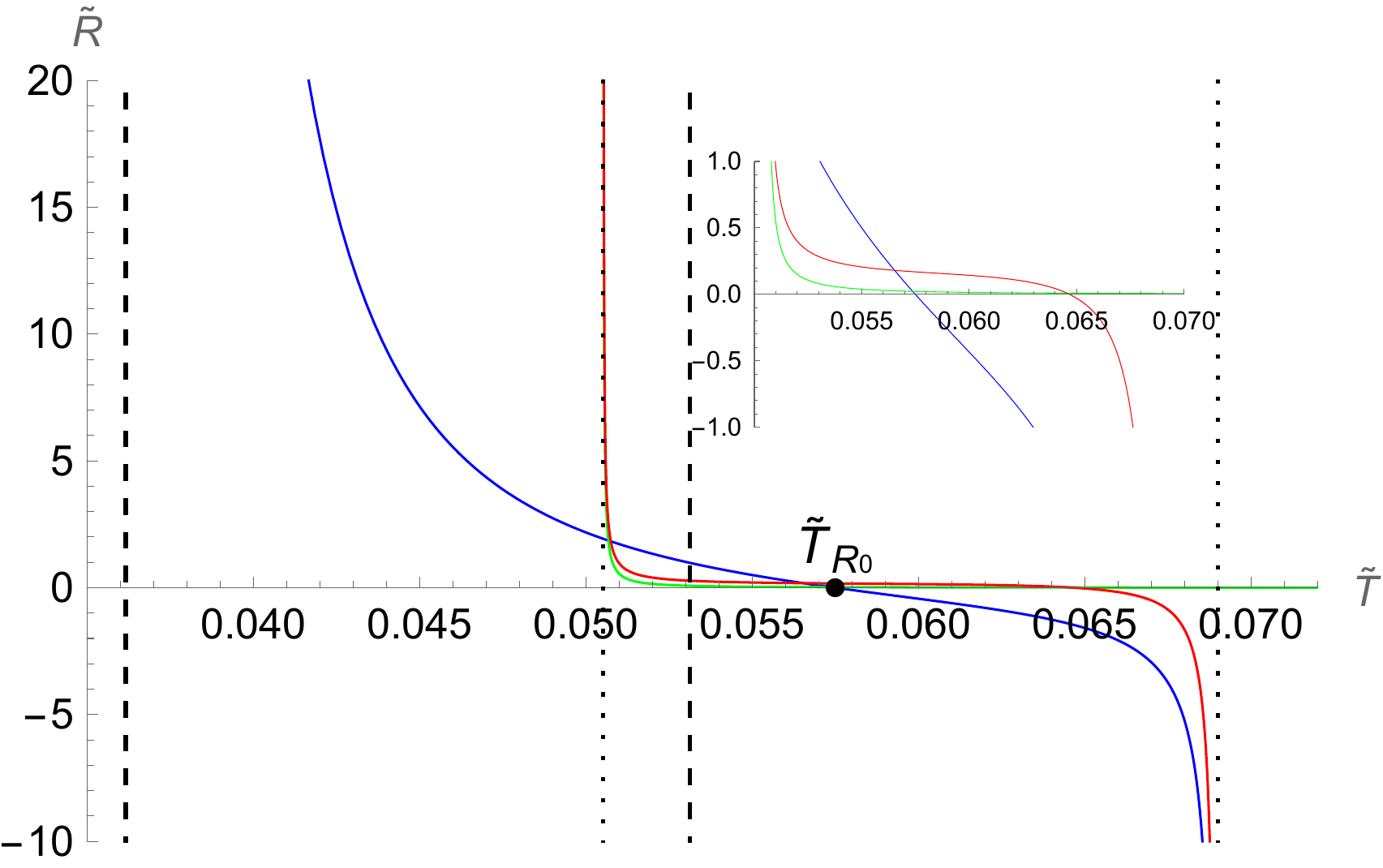}\label{fig:GRTCavS1:c}
\hspace{1cm}
\includegraphics[width=0.4\textwidth]{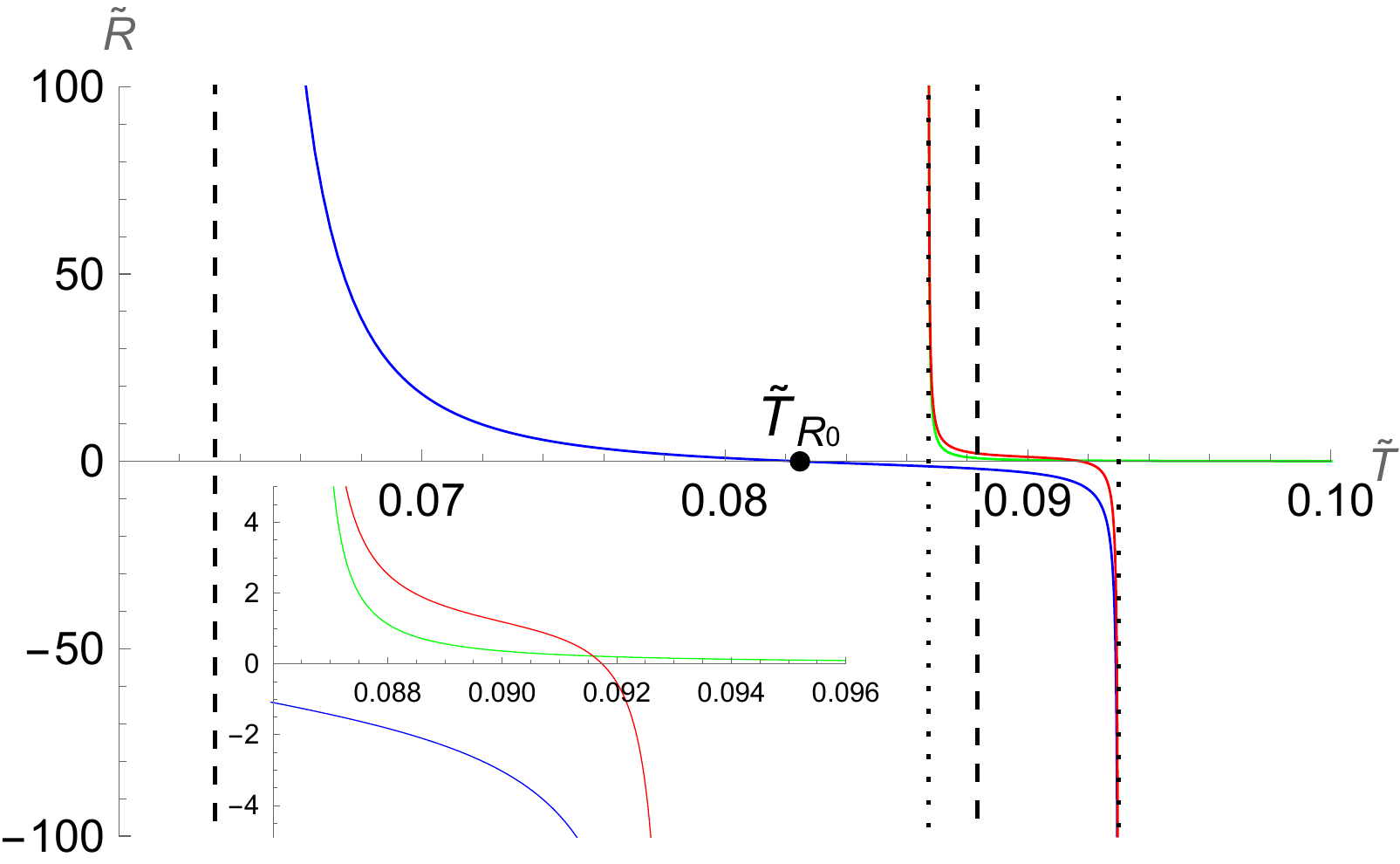}\label{fig:GRTCavS1:d}
\end{center}
\caption{Plots of the rescaled Gibbs free energy $\tilde{G}\ $and the rescaled
Ruppeiner invariant $\tilde{R}$\ against the rescaled temperature $\tilde{T}%
$\ for RN black holes in a cavity with the rescaled pressure $\tilde{P}=0.002$
(left column) and $0.006$ (right column). There are three black hole phases
for $\tilde{T}_{1}<\tilde{T}_{p}<\tilde{T}_{2}$, and a first-order transition
between Small BH and Large BH occurring at $\tilde{T}=\tilde{T}_{p}$. For
Small and Intermediate BHs, $\tilde{R}=-\infty$ at $\tilde{T}=\tilde{T}_{2}$.
The black points correspond to $\tilde{R}=0$ at $\tilde{T}=\tilde{T}_{R_{0}}$.
\textbf{Left Column}: $\tilde{T}_{R_{0}}>\tilde{T}_{p}$. The interactions of
the globally preferred phases are always repulsive. \textbf{Right Column:
}$\tilde{T}_{R_{0}}<\tilde{T}_{p}$. For the globally preferred phases, the
type of interactions changes from repulsive to attractive at $\tilde{T}%
=\tilde{T}_{R_{0}}$ and then returns to repulsive at $\tilde{T}=\tilde{T}_{p}$
as $\tilde{T}$ increases from $\tilde{T}_{\min}$.}%
\label{fig:GRTCavS1}%
\end{figure}\begin{figure}[tb]
\begin{center}
\includegraphics[width=0.4\textwidth]{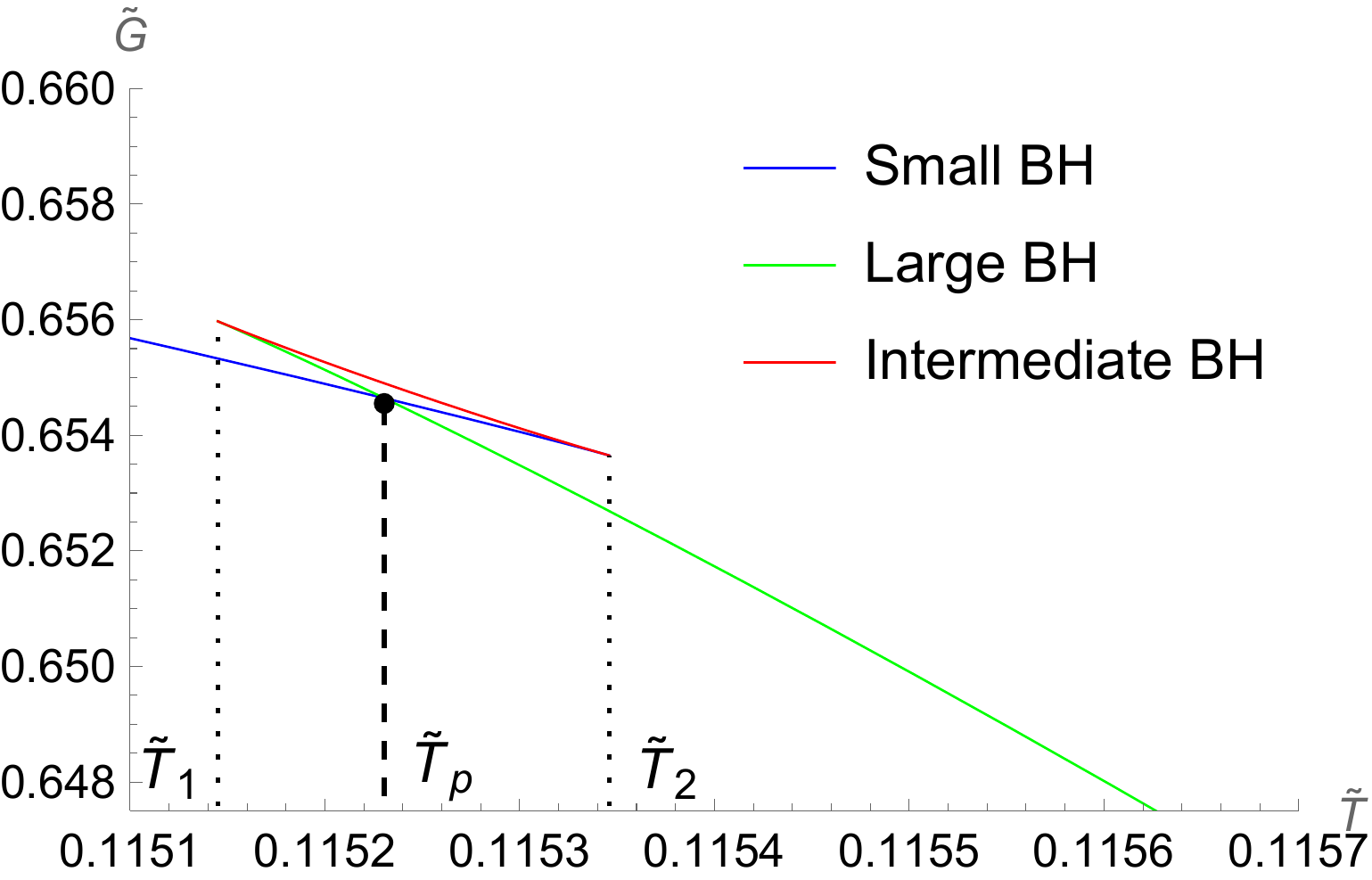}\label{fig:GRTCavS2:a}
\hspace{1cm}
\includegraphics[width=0.4\textwidth]{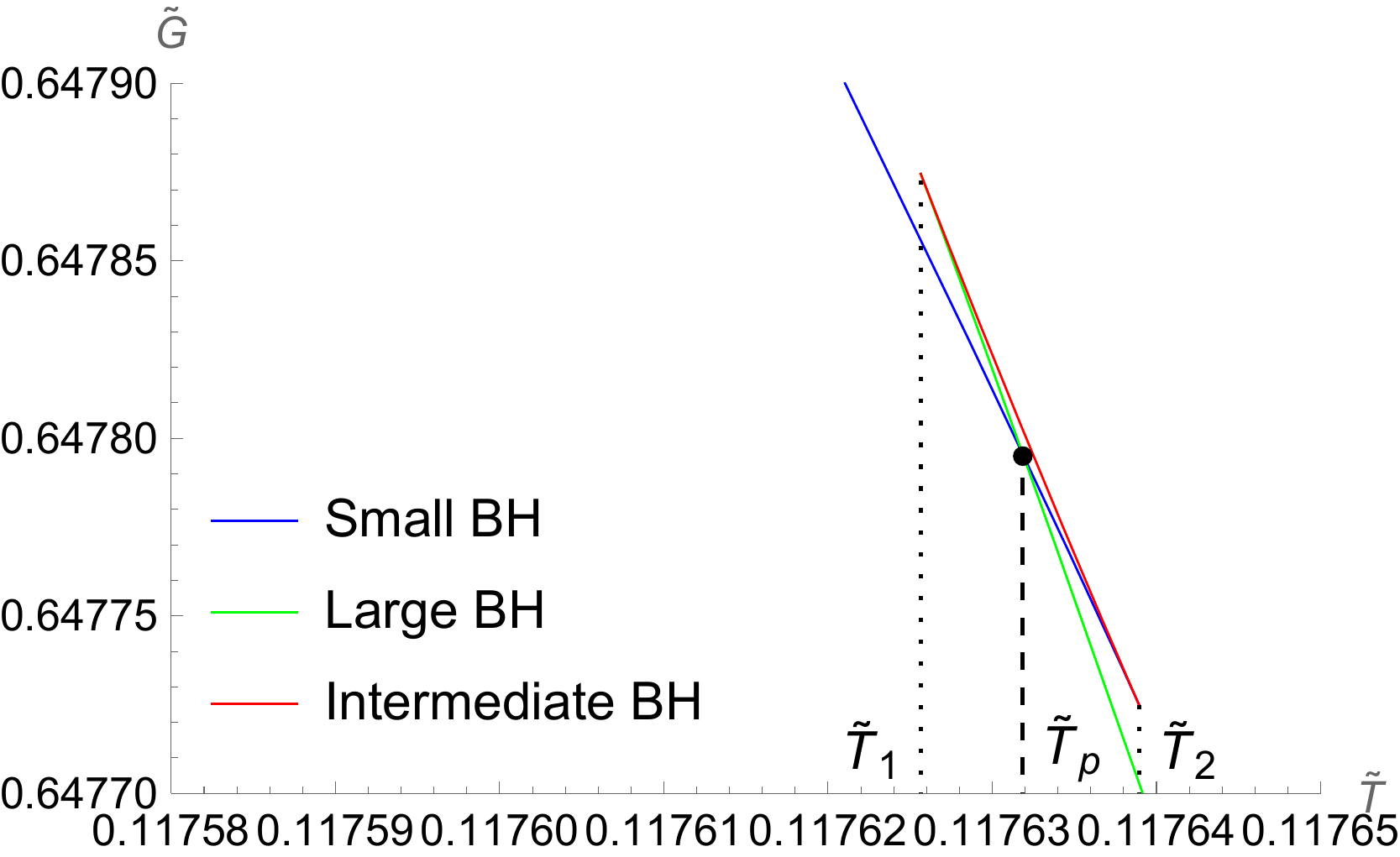}\label{fig:GRTCavS2:b}
\includegraphics[width=0.4\textwidth]{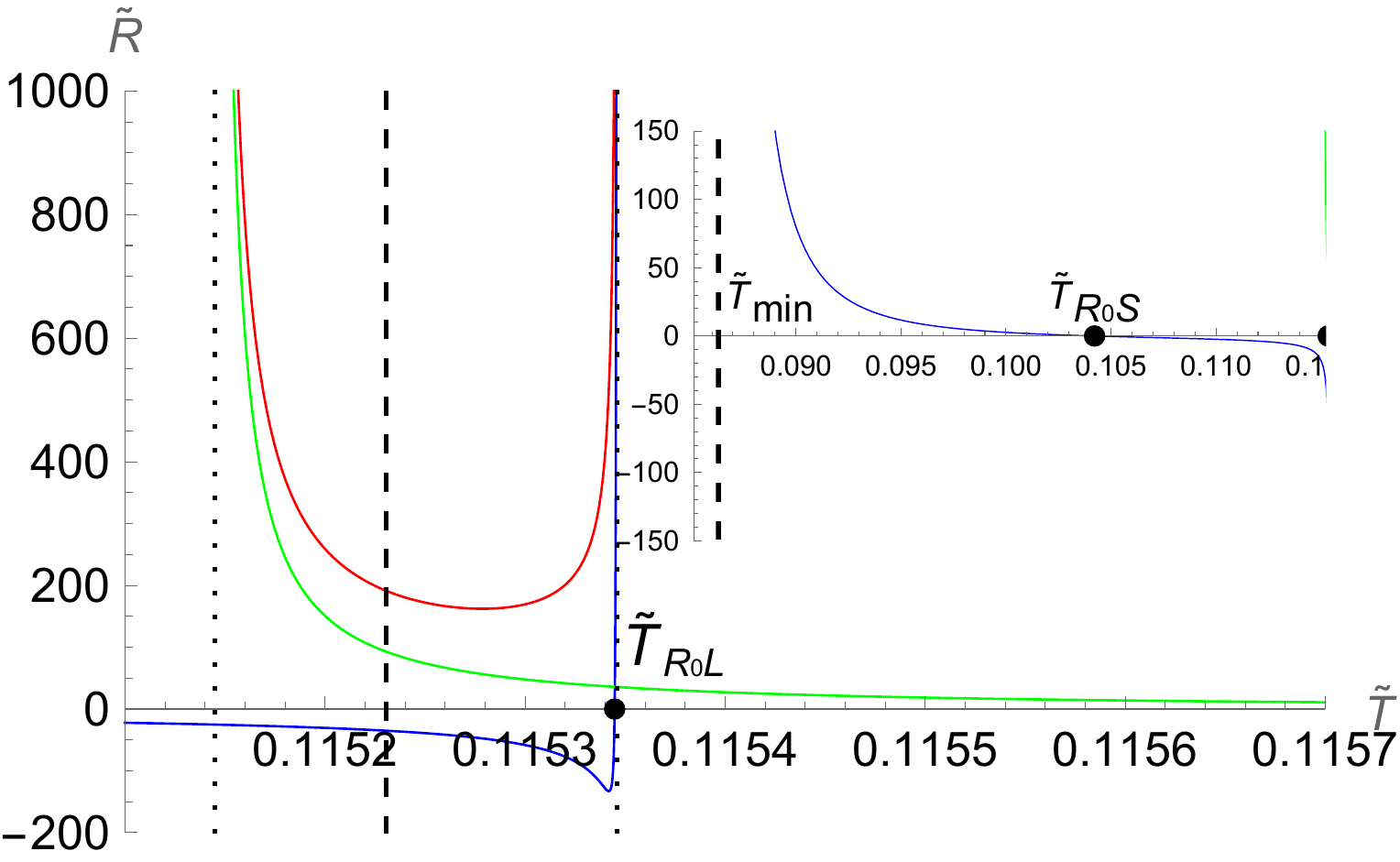}\label{fig:GRTCavS2:c}
\hspace{1cm}
\includegraphics[width=0.4\textwidth]{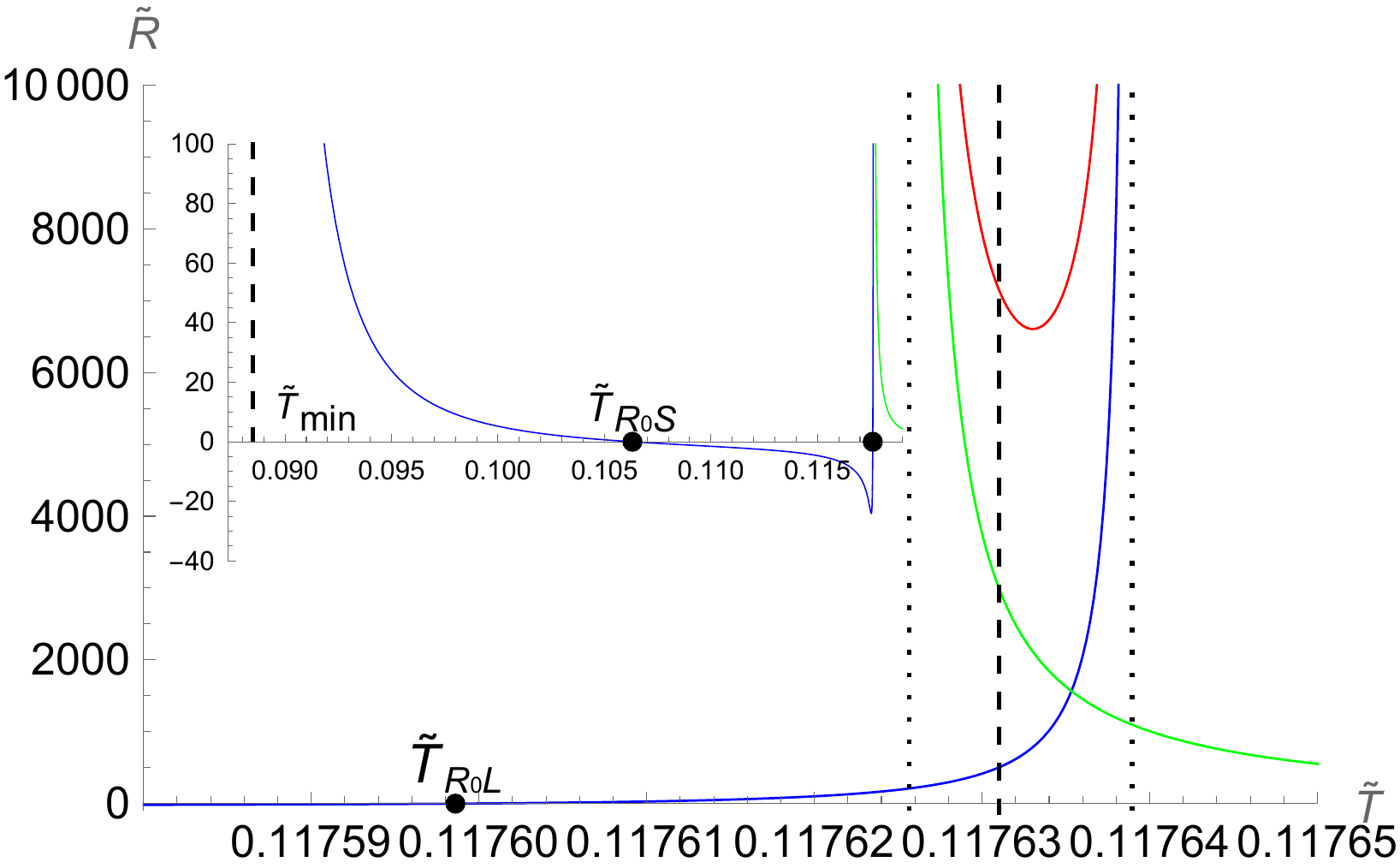}\label{fig:GRTCavS2:d}
\end{center}
\caption{Plots of the rescaled Gibbs free energy $\tilde{G}\ $and the rescaled
Ruppeiner invariant $\tilde{R}$\ against the rescaled temperature $\tilde{T}%
$\ for RN black holes in a cavity with the rescaled pressure $\tilde
{P}=0.0109$ (left column) and $0.0114$ (right column). Three black hole phases
coexist for $\tilde{T}_{1}<\tilde{T}_{p}<\tilde{T}_{2}$, during which a
first-order transition between Small BH and Large BH occurs at $\tilde
{T}=\tilde{T}_{p}$. The $\tilde{R}$ of Large and Intermediate BHs are
positive, whereas Small BH has a negative $\tilde{R}$ for $\tilde{T}_{R_{0}%
S}<\tilde{T}<\tilde{T}_{R_{0}L}$ and vanishing $\tilde{R}$ at $\tilde
{T}=\tilde{T}_{R_{0}S}$ and $\tilde{T}_{R_{0}L}$, marked by black points.
\textbf{Left Column}: $\tilde{T}_{R_{0}L}>\tilde{T}_{p}$. The microstructure
interactions of the globally preferred phases turn attractive from repulsive
at $\tilde{T}=\tilde{T}_{R_{0}S}$, and then return to repulsive at $\tilde
{T}=\tilde{T}_{p}$ as one increases $\tilde{T}$ from $\tilde{T}_{\min}$.
\textbf{Right Column:} $\tilde{T}_{p}>\tilde{T}_{R_{0}L}$. For the globally
preferred phases, the type of interactions changes from repulsive to
attractive at $\tilde{T}=\tilde{T}_{R_{0}S}$ and then returns to repulsive at
$\tilde{T}=\tilde{T}_{R_{0}L}$ as $\tilde{T}$ increases from $\tilde{T}_{\min
}$. However, the interactions remain repulsive during the phase transition at
$\tilde{T}=\tilde{T}_{p}$.}%
\label{fig:GRTCavS2}%
\end{figure}

Furthermore, we can obtain $\tilde{r}_{+}$ as a function of $\tilde{T}$ and
$\tilde{P}$ by solving eqn. $\left(  \ref{eq:CavityTP}\right)  $, and then
express $\tilde{G}$ and $\tilde{R}$ in terms of $\tilde{T}$ and $\tilde{P}$.
With $\tilde{G}(\tilde{T},\tilde{P})$ and $\tilde{R}(\tilde{T},\tilde{P})$, we
can discuss the phase structure of RN black holes in a cavity for given
$\tilde{T}$ and $\tilde{P}$, and compute the Ruppeiner invariant of each
phase. In the extended phase spaces, it has been shown that the phase
structure of RN black holes in a cavity is similar to that of AdS counterparts
\cite{Wang:2020hjw}. Specifically, FIGs. \ref{fig:GRTCavS1} and
\ref{fig:GRTCavS2} show that, when $\tilde{P}<\tilde{P}_{c}$, RN black holes
in a cavity have three phases, namely Small BH, Intermediate BH and Large BH,
coexisting for $\tilde{T}_{1}<\tilde{T}<\tilde{T}_{2}$, and there is a
first-order phase transition between Small BH and Large BH occurring at
$\tilde{T}=\tilde{T}_{p}$. However, the behavior of the Ruppeiner invariant is
quite richer than the AdS case. In fact, we find

\begin{itemize}
\item $\tilde{P}<\tilde{P}_{1}\simeq0.0108$. The $\tilde{R}$ of Small BH
decreases monotonically from positive infinity to negative infinity as one
increases $\tilde{T}$ from $\tilde{T}_{\min}$ to $\tilde{T}_{2}$. Therefore,
there exists $\tilde{R}=0$ at $\tilde{T}=\tilde{T}_{R_{0}}$, which could be
greater or smaller than $\tilde{T}_{p}$. We present $\tilde{G}$ and $\tilde
{R}$ as a function of $\tilde{T}$ for $\tilde{P}=0.002$ in the left column of
FIG. \ref{fig:GRTCavS1}, where $\tilde{T}_{R_{0}}>\tilde{T}_{p}$. Since
$\tilde{R}$ of Large BH is positive, $\tilde{R}$ stays positive at the phase
transition. So the globally preferred phases always have a positive $\tilde
{R}$, corresponding to repulsive interactions. On the other hand, we consider
the $\tilde{P}=0.006$ case in the right column of FIG. \ref{fig:GRTCavS1},
where $\tilde{T}_{R_{0}}<\tilde{T}_{p}$. Therefore, the globally preferred
phases have a negative $\tilde{R}$ when $\tilde{T}_{R_{0}}<\tilde{T}<\tilde
{T}_{p}$, which means that, as one increase $\tilde{T}$ from $\tilde{T}_{\min
}$, the interactions are repulsive at first, then become attractive at
$\tilde{T}=\tilde{T}_{R_{0}}$, and finally return to repulsive ones at
$\tilde{T}=\tilde{T}_{p}$.

\item $\tilde{P}_{1}<\tilde{P}<\tilde{P}_{c}$. The $\tilde{R}$ of Small BH is
positively infinite at $\tilde{T}=\tilde{T}_{2}$ and $\tilde{T}_{\min}$, and
has a negative minimum between $\tilde{T}_{2}$ and $\tilde{T}_{\min}$, which
means $\tilde{R}=0$ occurs at $\tilde{T}=\tilde{T}_{R_{0}S}$ and $\tilde
{T}_{R_{0}L}$, respectively, with $\tilde{T}_{\min}<\tilde{T}_{R_{0}S}%
<\tilde{T}_{R_{0}L}<\tilde{T}_{2}$. Although we find $\tilde{T}_{R_{0}S}$ is
always smaller than $\tilde{T}_{p}$, $\tilde{T}_{R_{0}L}$ may be smaller or
greater than $\tilde{T}_{p}$. Note that Large BH always has a positive
$\tilde{R}$. For $\tilde{P}=0.0109$, $\tilde{G}$ and $\tilde{R}$ are plotted
against $\tilde{T}$ in the left column of FIG. \ref{fig:GRTCavS2}, where
$\tilde{T}_{R_{0}L}>\tilde{T}_{p}$. In this case, the globally preferred
phases have a negative $\tilde{R}$ and hence attractive interactions for
$\tilde{T}_{R_{0}S}<\tilde{T}<\tilde{T}_{p}$. When $\tilde{T}>\tilde{T}_{p}$
or $\tilde{T}<\tilde{T}_{R_{0}S}$, the interactions of the globally preferred
phases are repulsive, which indicates that the type of the interactions
changes during the phase transition. We show the $\tilde{P}=0.0114$ case in
the right column of FIG. \ref{fig:GRTCavS2}, where $\tilde{T}_{R_{0}L}%
<\tilde{T}_{p}$. At $\tilde{T}=\tilde{T}_{p}$, Large and Small BHs both have a
positive $\tilde{R}$, indicating that the type of the interactions remains
repulsive during the phase transition. Moreover, as $\tilde{T}$ increases from
$\tilde{T}_{\min}$ to $\tilde{T}_{p}$, the globally preferred phase (i.e.,
Small BH) undergoes repulsive microstructure interactions $\rightarrow$
attractive ones $\rightarrow$ repulsive ones.\begin{figure}[tb]
\begin{center}
\includegraphics[width=0.35\textwidth]{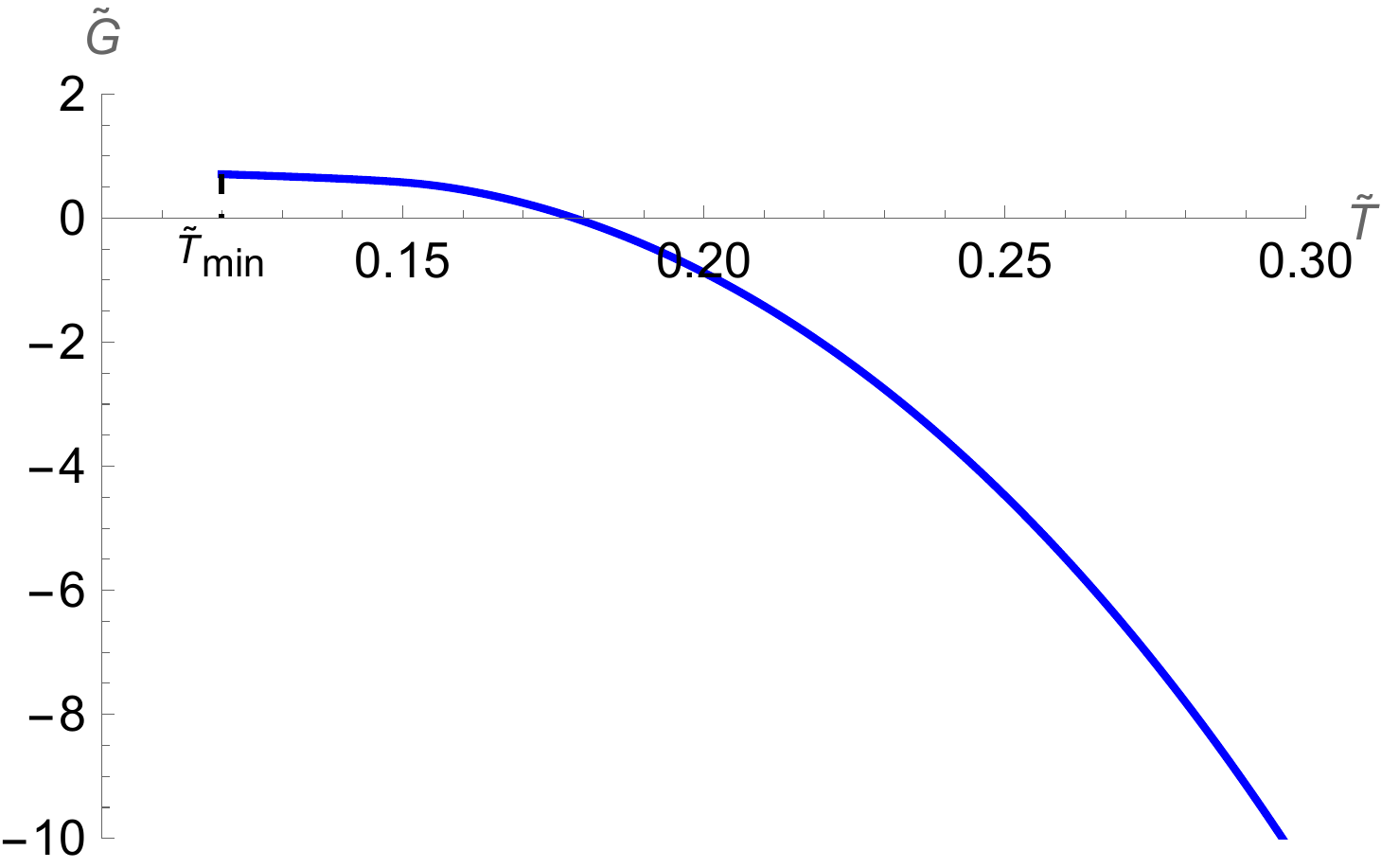}\label{fig:GRTCavL:a}
\hspace{1cm}
\includegraphics[width=0.35\textwidth]{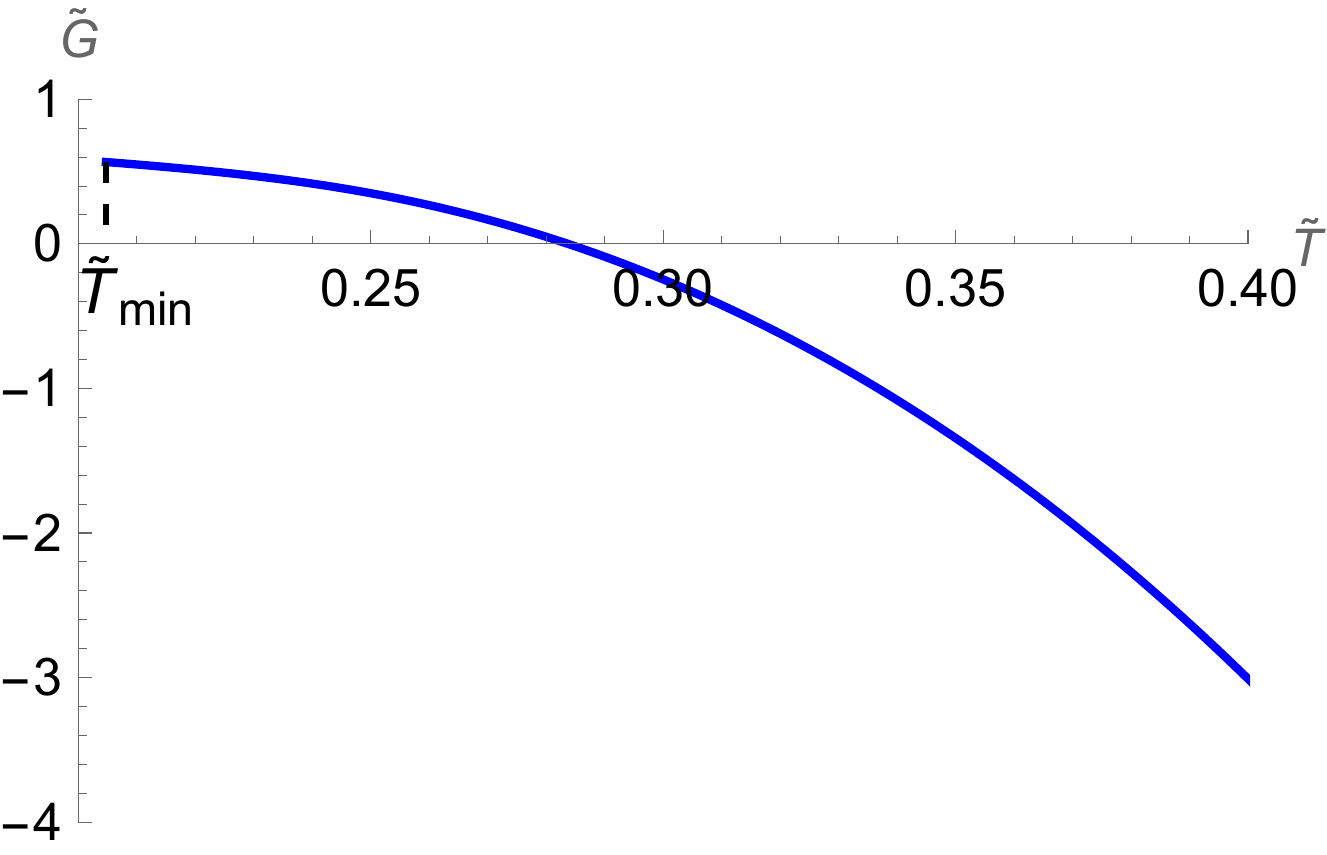}\label{fig:GRTCavL:b}
\includegraphics[width=0.35\textwidth]{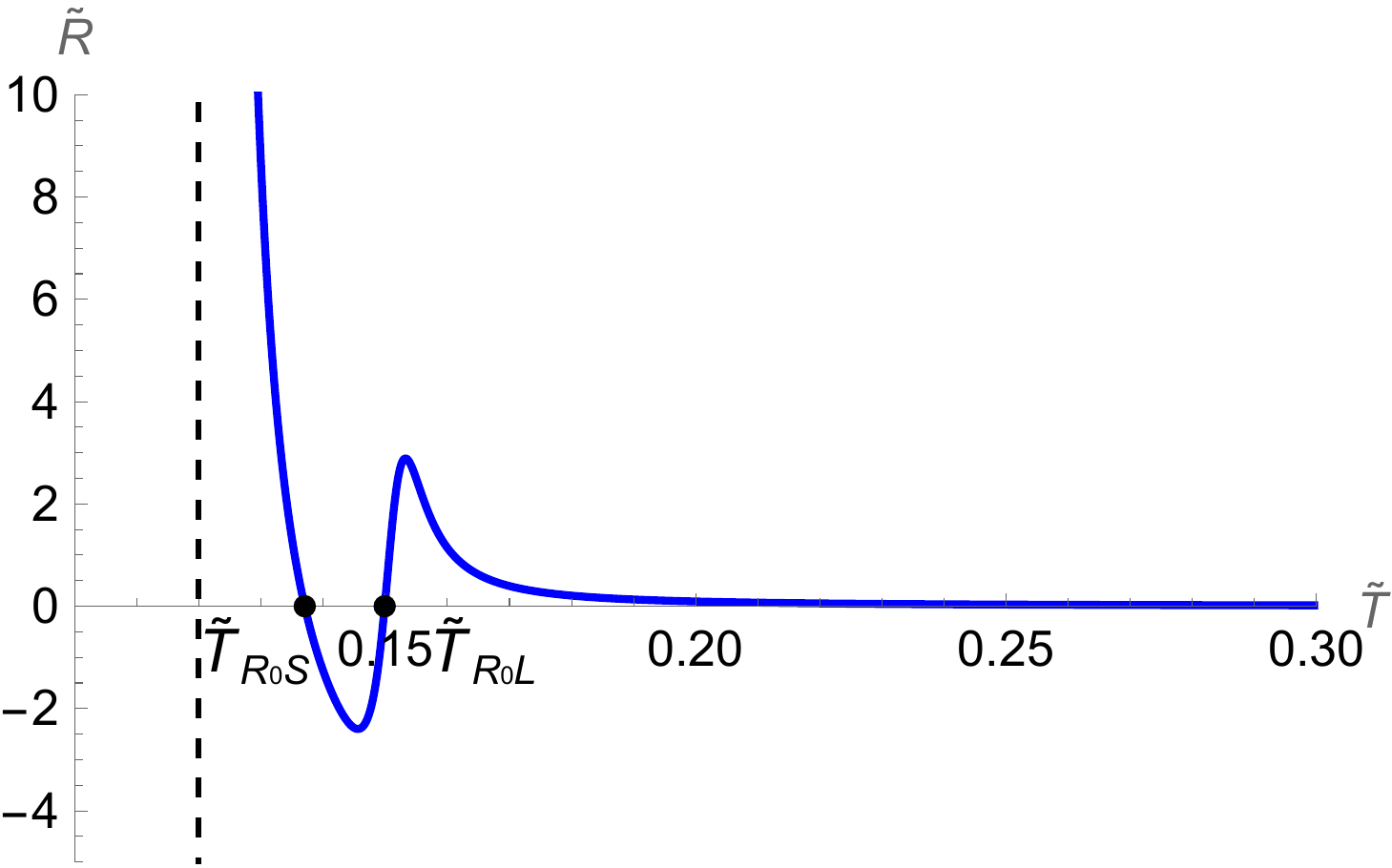}\label{fig:GRTCavL:c}
\hspace{1cm}
\includegraphics[width=0.35\textwidth]{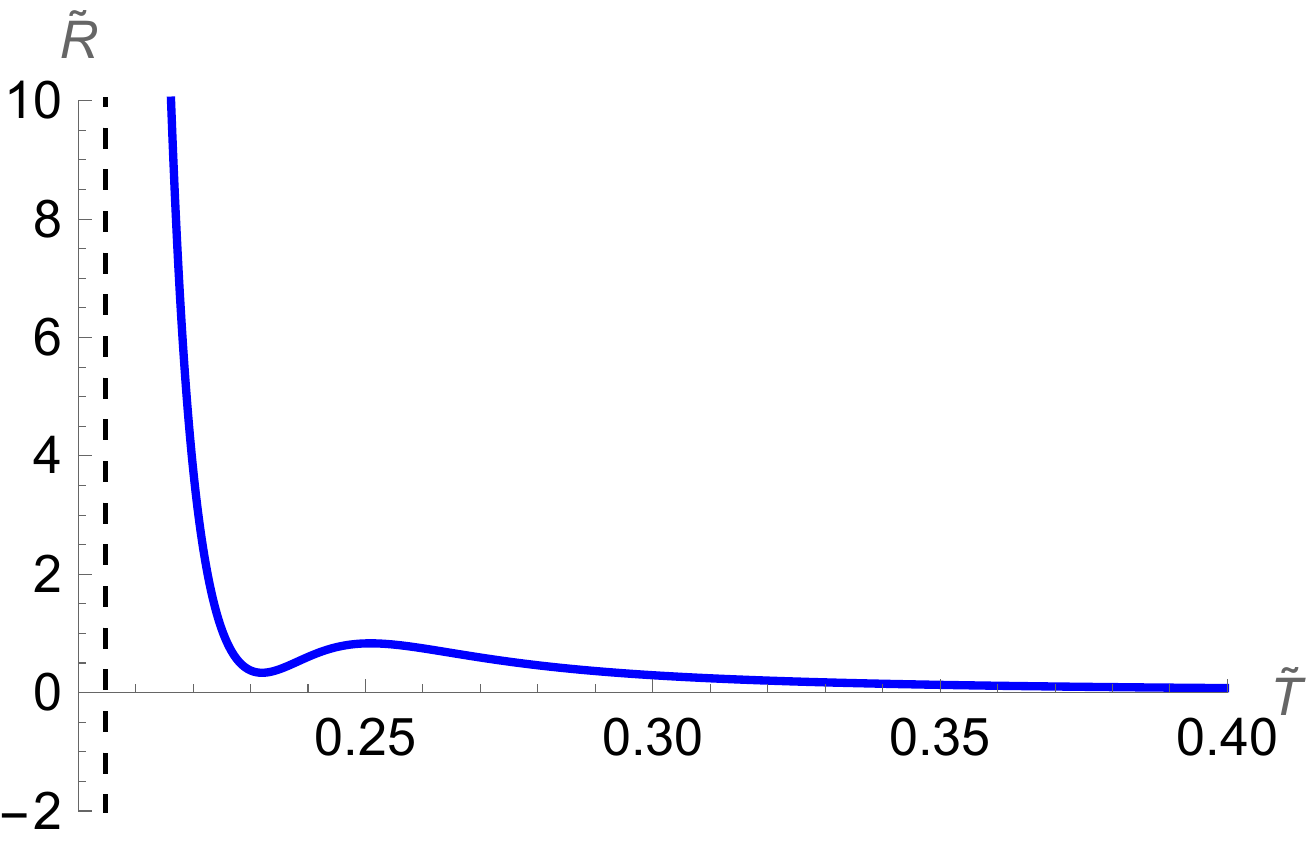}\label{fig:GRTCavL:d}
\end{center}
\caption{Plots of the rescaled Gibbs free energy $\tilde{G}\ $and the rescaled
Ruppeiner invariant $\tilde{R}$\ against the rescaled temperature $\tilde{T}%
$\ for RN black holes in a cavity when $\tilde{P}>\tilde{P}_{c}$. Since there
is only one phase, no phase transitions occur. \textbf{Left Column}:
$\tilde{P}=0.02$. There are two zeros of $\tilde{R}=0$, which are $\tilde
{T}_{R_{0}S}$ and $\tilde{T}_{R_{0}L}$, marked by black points. The
interactions turn attractive from repulsive, and then return to repulsive as
one increases $\tilde{T}$ from $\tilde{T}_{\min}$. \textbf{Right Column}:
$\tilde{P}=0.05$. The microstructure interactions are always repulsive.}%
\label{fig:GRTCavL}%
\end{figure}
\end{itemize}

When $\tilde{P}>\tilde{P}_{c}$, there is only one phase and no phase
transition. We find that $\tilde{R}$ always has a maximum and a minimum, and
approaches positive infinity and zero as $\tilde{T}\rightarrow\tilde{T}_{\min
}$ and $\tilde{T}\rightarrow\infty$, respectively. In the left column of FIG.
\ref{fig:GRTCavL}, we plot $\tilde{G}$ and $\tilde{R}$ as a function of
$\tilde{T}$ for $\tilde{P}=0.02$. In this case, $\tilde{R}$ has two zeros at
$\tilde{T}=\tilde{T}_{R_{0}S}$ and $\tilde{T}_{R_{0}L}$, and hence the
microstructure interactions are attractive for $\tilde{T}_{R_{0}S}<\tilde
{T}<\tilde{T}_{R_{0}L}$ and repulsive otherwise. The $\tilde{P}=0.05$ case is
plotted in the right column of FIG. \ref{fig:GRTCavL}, which shows that
$\tilde{R}$ is always positive, and therefore the microstructure interactions
are always repulsive.

\begin{figure}[tb]
\begin{center}
\includegraphics[width=0.4\textwidth]{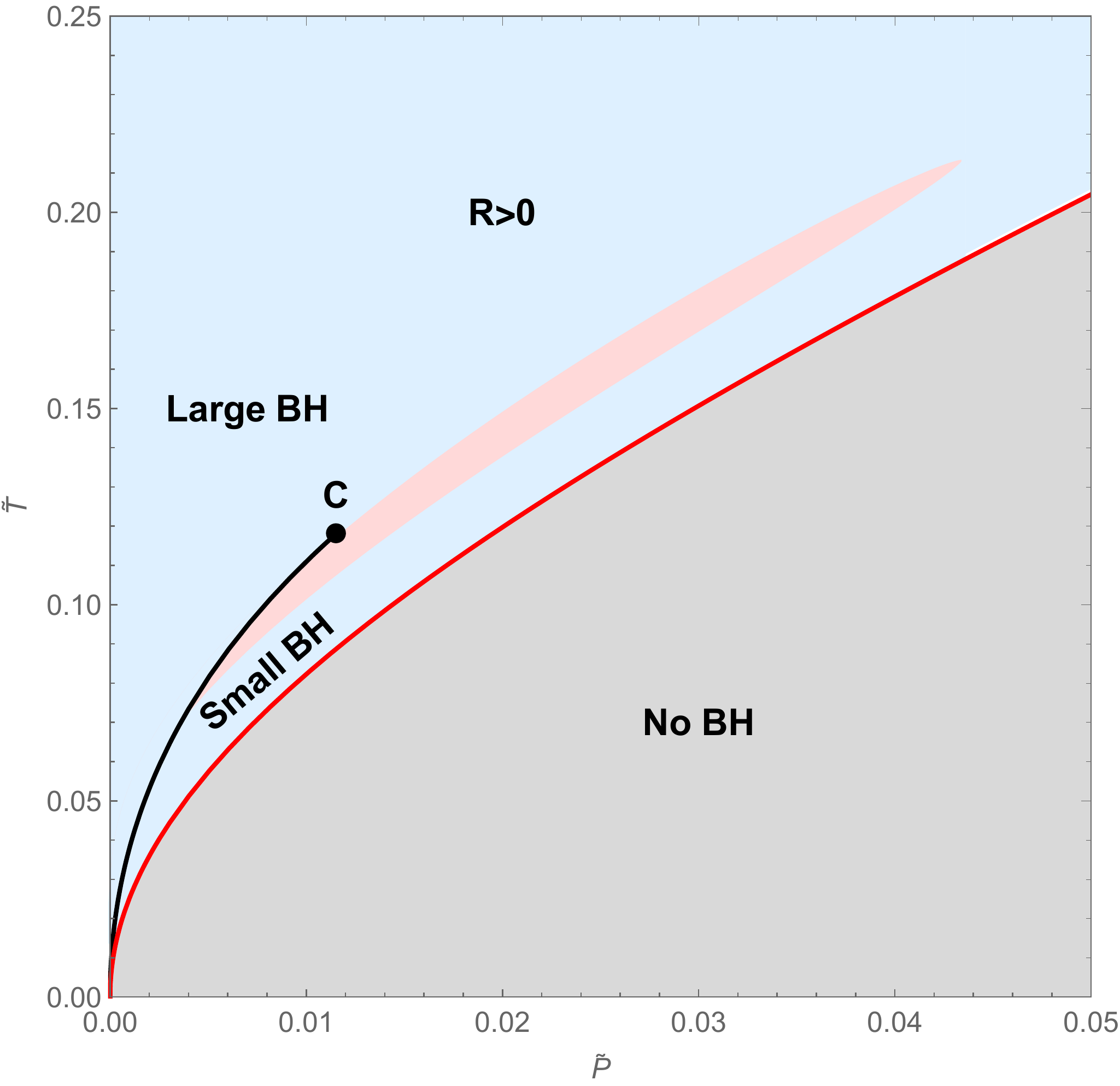} \hspace{1cm}
\includegraphics[width=0.41\textwidth]{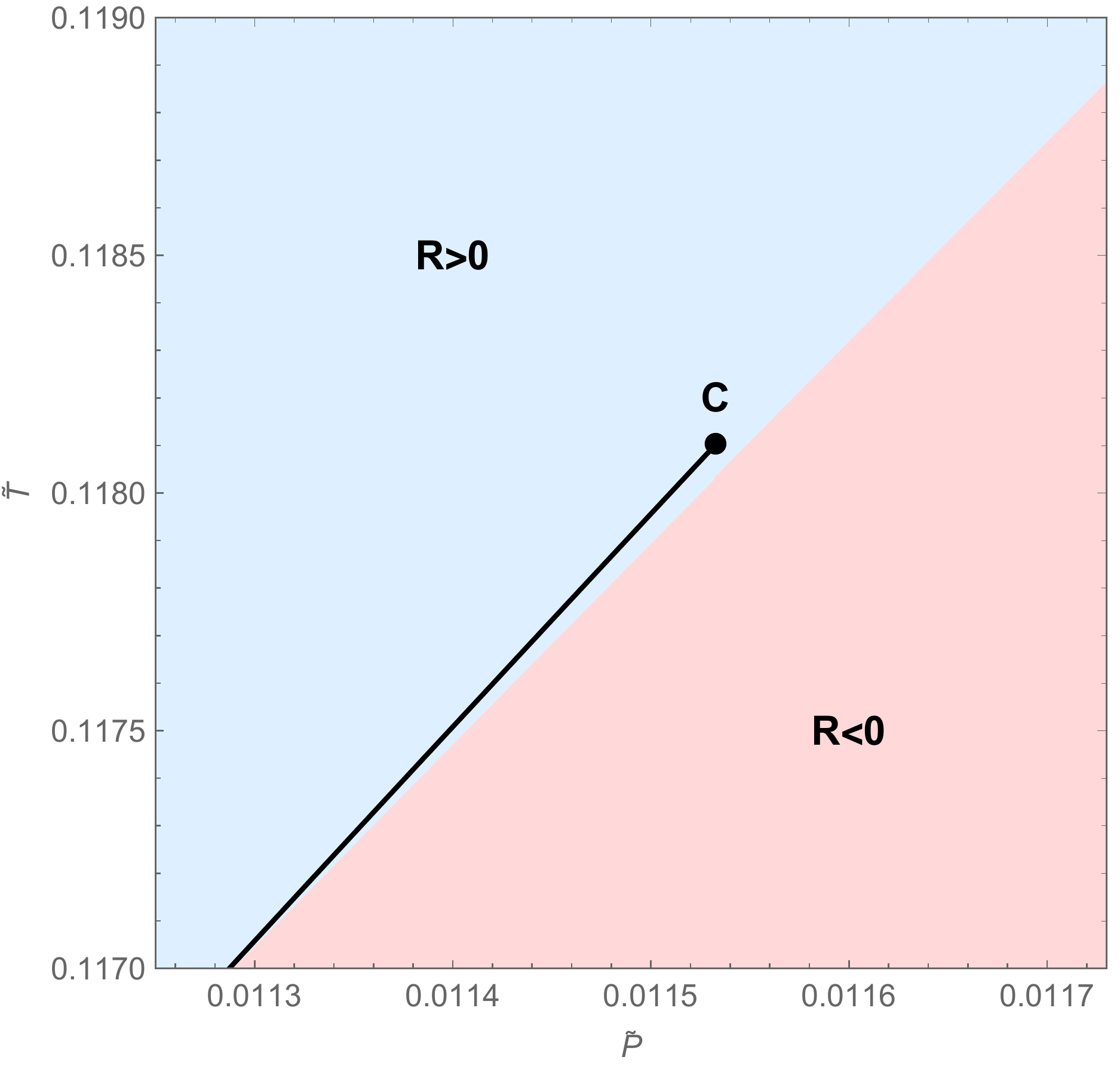}
\end{center}
\caption{Phase diagram of RN black holes in a cavity in the $\tilde{P}%
$-$\tilde{T}$ plane. The black lines and black dots represent the first-order
phase transition and the critical point, respectively. In the blue (red)
region, $R>0$ ($R<0$), corresponding to repulsive (attractive) interactions.
Black hole solutions do not exist in the gray region. On the red line, one has
$R=+\infty$.}%
\label{fig:TPR0Cav}%
\end{figure}

In FIG. \ref{fig:TPR0Cav}, we present the globally preferred phases of RN
black holes in a cavity in the $\tilde{P}$-$\tilde{T}$ plane. The first-order
phase transition between Large BH and Small BH is represented by a black line,
and the second-order critical point is marked by a black dot. The phase
diagram of RN black holes in a cavity is quite similar to that of RN-AdS black
holes, except for the gray region where no black hole solutions exist. The
region where $\tilde{R}>0$ ($\tilde{R}<0$) is displayed as the blue (red)
region. It shows that the microstructure interactions are repulsive in most
part of the parametric space, while the $\tilde{R}<0$ region is an
\textquotedblleft island\textquotedblright\ in the $\tilde{P}$-$\tilde{T}$
plane. The right panel of FIG. \ref{fig:TPR0Cav} highlights the region near
the critical point, and exhibits that the interactions around the critical
point are repulsive. Unlike RN-AdS black holes, the globally preferred phases
of RN black holes in a cavity can have a divergent $\tilde{R}$ on the red
line, corresponding to black holes at $\tilde{T}=\tilde{T}_{\min}$.

\section{Discussions and Conclusions}

In this paper, we studied the phase structure and the thermodynamic geometry
of charged (RN) black holes in the extended phase space by considering two
kinds of boundary conditions, namely the asymptotically AdS boundary and the
Dirichlet boundary in the asymptotically flat spacetime. For the extended
phase space of RN-AdS black holes, the cosmological constant is interpreted as
a pressure, and the conjugate thermodynamic volume then is $V=$ $4\pi
r_{+}^{3}/3$ with $r_{+}$ being the horizon radius. In parallel with AdS black
holes, we extended the phase space of RN black holes in a cavity by
introducing a thermodynamic volume $V=4\pi r_{B}^{3}/3$, where $r_{B}$ is the
radius of the cavity, and the conjugate pressure $P=-\partial E/\partial V$,
where $E$ is the thermal energy. In these extended phase spaces, we studied
the thermodynamic geometry of charged black holes by taking the internal
energy $U$ and the volume $V$ as fluctuation variables. Although the phase
structures of the AdS and cavity cases were shown to be alike, we found that
there are significant differences between the thermodynamic geometries in
these two cases.

Specifically, we calculated the normalized Ruppeiner invariant $\bar{R}\equiv
RC_{V}$ for RN-AdS black holes and the rescaled Ruppeiner invariant $\tilde
{R}=RQ^{2}$ for RN black holes in a cavity. The behavior of $\bar{R}$ and
$\tilde{R}$ as a function of the rescaled horizon radius $\tilde{r}_{+}$ and
the rescaled pressure $\tilde{P}$ were presented in FIGs. \ref{fig:rPR0AdS}
and \ref{fig:rPR0Cav}, respectively. In the AdS case, $\bar{R}$ only diverges
to positive infinity on the red line, and goes to $-1/2$ in the extremal
limit. In the cavity case, $\tilde{R}$ can diverge to negative infinity or
positive infinity, and goes to positive infinity as the temperature approaches
its minimum value. We displayed phase diagrams in the $\tilde{T}$-$\tilde{P}$
plane of RN-AdS black holes and RN black holes in a cavity, as well as the
parametric regions where $R>0$ or $R<0$, in FIGs. \ref{fig:TPR0AdS} and
\ref{fig:TPR0Cav}, respectively. For RN-AdS black holes with a fixed
$\tilde{P}$, $\bar{R}=0$ always has a single solution, corresponding to the
change of the type of the microstructure interactions as $\tilde{T}$ increases
at a constant $\tilde{P}$. For RN black holes in a cavity with a given
$\tilde{P}$, $\bar{R}=0$ has either no solution, indicating the interactions
always stay repulsive as $\tilde{T}$ increases at a constant $\tilde{P}$, or
exactly two solutions, corresponding to a reentrant transition of the type of
the interactions (i.e., repulsive $\rightarrow$ attractive $\rightarrow$
repulsive) as $\tilde{T}$ increases at a constant $\tilde{P}$.

In existing studies, the thermodynamic geometry of RN\ black holes in a cavity
was investigated in the normal phase space, in which the cavity radius is
fixed \cite{Wang:2019cax}. It was found that the microstructure interactions
always stay attractive before and after the LBH/SBH first-order transition.
The behavior of the thermodynamic geometry has been shown to be richer in the
extended phase space. On the other hand, black holes in a cavity have been
observed to possess a lot of interesting thermodynamic behavior, e.g., the
second law of thermodynamics and reentrant phase transitions, in the normal
phase space \cite{Wang:2019kxp,Liang:2019dni,Wang:2020osg}. It would be very
inspiring to explore thermodynamic phenomena other than the thermodynamic
geometry\ for black holes in a cavity in the extended phase space.

\label{Sec:DC}

\begin{acknowledgments}
We are grateful to Guangzhou Guo and Qingyu Gan for useful discussions and
valuable comments. This work is supported in part by NSFC (Grant No. 11875196,
11375121, 11947225 and 11005016).
\end{acknowledgments}

\bibliographystyle{unsrturl}
\bibliography{ref}

\end{document}